\documentclass[12pt]{article}

\usepackage{epsfig}
 
\usepackage{amsfonts}
\usepackage{amscd}
\usepackage{latexsym}
\usepackage{amsmath,amssymb}
\usepackage{verbatim}
\usepackage{color}
\usepackage{cite}
\usepackage[textheight=9in, textwidth=6.5in, letterpaper]{geometry}
\usepackage{soul}
\usepackage{appendix}

\usepackage{hyperref}

\def\az{{\bf A}_0}
\usepackage{graphicx}
\usepackage{tikz}
\usepackage{tikz-3dplot}
\usetikzlibrary{calc}
\usetikzlibrary{positioning}
\usepackage{float}
\usepackage{subfig}
\usepackage[T1]{fontenc}
\usepackage{amsmath, amssymb, amsthm, fullpage}

\usepackage{doc}
\usepackage{ifthen, color, fancyvrb}
\usepackage{nextpage}

\usepackage{wrapfig}
\usepackage{parskip}
\usepackage{authblk}
\usepackage[makeroom]{cancel}
\usepackage{slashed}
\def\be{\begin{equation}}
\def\ee{\end{equation}}
\def\bz{{\bar z}}
\def\bw{{\bar w}}
\def\p{\partial}
\def\ci{{\cal I}}

\def\be{\begin{equation}}
\def\ee{\end{equation}}
\def\bea{\begin{eqnarray}}
\def\eea{\end{eqnarray}}
\def\<{\langle }
\def\>{\rangle}

\def\l{{\lambda}}
\def\D{{\Delta}}
\def\ve{{\varepsilon}}
\def\eps{{\epsilon}}
\def\bh{\bar{h}}
\def\om{\omega}
\def\ca{\mathcal{A}}

\def\cs{\mathcal{S}}
\def\co{{\mathcal O}}

\def\g{\gamma }
\def\ms{{\bf M} }

\begin{document}
\begin{titlepage}
\unitlength = 1mm
\ \\
\vskip 2cm
\begin{center}
{\LARGE{\textsc{Celestial Amplitudes from UV to IR}}}

\vspace{1cm}
Nima Arkani-Hamed$^1$, Monica Pate$^{2,3}$, Ana-Maria Raclariu$^{2,4}$ and Andrew Strominger$^2$\\ \vspace{2ex}
\noindent$^1${\it The Institute for Advanced Study, Princeton, NJ, USA}\\
\noindent $^2${\it  Center for the Fundamental Laws of Nature, Harvard University,
Cambridge, MA, USA}\\
\noindent$^3${\it Society of Fellows, Harvard University,
Cambridge, MA, USA}\\
\noindent$^4${\it Perimeter Institute for Theoretical Physics, Waterloo, ON, Canada}\\

\vspace{0.8cm}

\begin{abstract}

Celestial amplitudes represent 4D scattering of particles in boost, rather than the usual energy-momentum, eigenstates and hence are sensitive to  both UV and IR physics.   We show that known UV and IR properties of quantum gravity  translate into powerful constraints on the analytic structure of celestial amplitudes. For example the soft UV behavior of quantum gravity is shown to imply that the exact four-particle scattering amplitude is meromorphic in the complex boost weight plane with poles confined to even integers on the negative  real axis.  Would-be poles on the positive real axis from UV asymptotics are shown to be erased by a flat space analog of the AdS resolution of the bulk point singularity. The residues of the poles on the negative axis are identified with operator coefficients in the IR effective action. Far along the real positive axis, the scattering is argued  to grow exponentially according to the black hole area law. Exclusive amplitudes are shown to simply  factorize into conformally hard and conformally soft factors. The soft factor contains all IR divergences and is given  by a celestial current algebra correlator of Goldstone bosons from spontaneously broken asymptotic symmetries. The hard factor describes the  scattering of hard particles together with the boost-eigenstate clouds of soft photons or gravitons required by asymptotic symmetries. These provide  an IR safe $\cal S$-matrix for the scattering of hard particles.

 \end{abstract}

\end{center}
\vspace{1cm}

\end{titlepage}

\pagestyle{empty}
\pagestyle{plain}

\def\vx{{\vec x}}
\def\p{\partial}
\def\po{$\cal P_O$}

\pagenumbering{arabic}


\tableofcontents

\section{Introduction}
Scattering amplitudes are amongst the most basic observables in fundamental physics. They are of immense experimental interest in high-energy physics, but are also of central theoretical importance as the only known observables of quantum gravity in asymptotically flat space. It is thus natural to look for a holographic theory that determines the $\cal S$-matrix, formulated only in terms of the on-shell particles propagating to the boundary at infinity, without making any reference to bulk spacetime evolution.

This presents a new set of challenges beyond what we have become accustomed to in thinking about holography in AdS spacetimes. The boundary of AdS is an ordinary Lorentzian spacetime, with usual notions of locality and time evolution, and thus a local quantum field theory on this boundary is a natural candidate for the holographic theory whose quantum states are directly identified with those of the bulk gravitational theory. We do not have this luxury in asymptotically flat space. We can label the asymptopia associated with the scattering process in many different ways. If we are working with massless particles in four dimensions, this data can be given in terms of null momenta, or spinor-helicity variables, or in terms of twistor or momentum-twistor variables, or in terms of Mandelstam invariants formed between pairs of momenta. However there is no obvious notion of ``locality'' and ``time evolution'' in any of these spaces, and so no obvious candidate for a holographic theory using familiar physical principles.

Relatedly, even absent a candidate for a holographic theory in AdS, if we were handed a set of boundary correlators, we would understand the rules for checking their consistency as implied by  conformal field theory.  It is striking that over 60 years after the analogous questions were posed by $\cal S$-matrix theorists in the 1960's, we still don't know the answer to the analogous question for flat space scattering amplitudes. Indeed the $\cal S$-matrix program of the 1960's was ultimately doomed by the inability to precisely characterize the way in which bulk causality is imprinted in the analytic structure of the $\cal S$-matrix.

By contrast the explosion of progress in the understanding of scattering amplitudes over the past few decades suggests a more adventurous approach to these mysteries, inviting us to look for new physical principles and mathematical structures that will directly determine the $\cal S$-matrix. The emergence of a unitary, appropriately local and casual quantum-mechanical/spacetime description from the $\mathcal{S}$-matrix would then be seen to follow as an output, rather than being taken as primary.

An especially natural geometric space to look for this new holographic description is the celestial sphere at null infinity \cite{deBoer:2003vf, Pasterski:2016qvg}. The kinematical connection to the variables describing ordinary momentum-space amplitudes is very simple in four dimensions. Null four-momenta $p_{\alpha \dot{\alpha}}$ are written in terms of spinor-helicity variables, $p_{\alpha \dot{\alpha}} = \lambda_{\alpha} \tilde{\lambda}_{\dot{\alpha}}$. For real momenta in Minkowski space, we have $\tilde{\lambda}_{\dot{\alpha}} = \eta (\lambda_\alpha)^*$, where $\eta=\pm$ corresponds to outgoing/incoming particles. Lorentz transformations act as SL$(2,\mathbb{C})$ on the $\alpha, \dot{\alpha}$ indices. It is natural to write
\begin{equation} \label{sphelvar}
\lambda_\alpha = \sqrt{\omega} \left(\begin{array}{c} z \\ 1 \end{array} \right), \quad \quad \tilde{\lambda}_{\dot{\alpha}} =\eta \sqrt{\omega} \left(\begin{array}{c} \bar{z} \\ 1 \end{array} \right).
\end{equation}
The variable $z$ gives the direction of the null momentum and specifies a point on the (Riemann) celestial sphere, while $\omega$ is real and specifies the energy. The use of spinor-helicity variables makes the action of the little group on scattering amplitudes manifest. Note that the null-momentum $p_{\alpha \dot {\alpha}}$ does not uniquely specify $\lambda, \tilde{\lambda}$, since we can rephase $\lambda \to e^{i \theta} \lambda$ and $\tilde{\lambda} \to e^{-i \theta} \tilde{\lambda}$.  In fixing the above form for $\lambda, \tilde{\lambda}$, we have made a particular choice for associating $\lambda, \tilde{\lambda}$ with $p_{\alpha \dot{\alpha}}$, where $\omega$ is taken to be real. The SL$(2,\mathbb{C})$ Lorentz transformations act nicely on $z$ and $\omega$ while giving us the appropriate little group rephasing: 
\begin{equation}
\sqrt{\omega} \left(\begin{array}{c} z \\ 1 \end{array} \right) \to \left(\begin{array}{cc} a & b \\ c & d \end{array}\right) \sqrt{\omega} \left(\begin{array}{c} z \\ 1 \end{array} \right) = e^{i \theta} \sqrt{\omega^\prime} \left(\begin{array}{c} z^\prime \\ 1 \end{array} \right),
\end{equation}
where
\begin{equation}
\omega^\prime = |cz + d|^2 \omega, \quad \quad z^\prime = \frac{a z + b}{cz + d}, \quad\quad e^{2 i \theta} = \frac{c z + d}{\bar{c} \bar{z} + \bar{d}}.
\end{equation}

In this way, any $n$ particle scattering amplitude ${\bf A}(\lambda_i, \tilde{\lambda}_i) = {\bf A}(\omega_i,z_i)$ is labeled by energies $\omega_i$ and points $z_i$ on the celestial sphere. Since the action of the Lorentz SL$(2,\mathbb{C})$ is simply M{\"o}bius transformations on $z$, it is natural to think of the amplitudes as transforming just like correlation functions of a conformal field theory on the celestial sphere. On the other hand, the usual momentum eigenstates don't transform nicely under conformal transformations. Instead, it is natural to scatter conformal primary states $|\Delta, z \rangle$, defined (for massless scalars) as \cite{Pasterski:2016qvg,Pasterski:2017kqt}
\begin{equation}
|\Delta, z \rangle \equiv \int_0^\infty \frac{d \omega}{\omega} \omega^{\Delta} |\omega,z\rangle.
\end{equation}
Then under the Lorentz SL$(2,\mathbb{C})$, we have $|\Delta, z \rangle \to |cz + d|^{-2\Delta}  |\Delta, \frac{a z + b}{cz + d}\rangle $. This is just the transformation of conformal primaries of dimension $\Delta$. Given that boosts in the direction of the null momentum leave $z$ invariant, we can also think of the state $|\Delta,z \rangle$ as an eigenstate of boosts (or Rindler energies) in the direction of $z$.

Thus the scattering processes most natural to the celestial sphere are not of momentum eigenstates, but boost eigenstates, related to the more familiar momentum-space amplitudes via a Mellin transform:
\begin{equation}
{\cal A}(\Delta_j,z_j) = \left(\prod_{i=1}^n\int_0^\infty \frac{d \omega_i}{\omega_i} \omega_i^{\Delta_i} \right) {\bf A}(\om_j,z_j) . 
\end{equation}
We will refer to these conformal basis scattering amplitudes as {\it celestial amplitudes}. 
Note that while in this way we have made the action of the Lorentz group manifest, translational invariance is not manifest and implies further restrictions on ${\cal A}(\Delta_i,z_i)$. For instance in the case of four-particle scattering, SL$(2,\mathbb{C})$ invariance tells us that non-trivial dependence on $z_{1}, \cdots, z_4$ can only be via the cross-ratio \be \label{crossratio} z \equiv  \frac{(z_1 - z_3)(z_2 - z_4)}{(z_1 - z_2)(z_3 - z_4)},\ee but translational invariance tells us more. By momentum conservation, the spatial momenta for all four particles must lie on a two-dimensional plane; this plane intersects the celestial sphere on a circle, and so the four $z_{1},\cdots,z_4$ must lie on that circle. The condition for four points to be co-circular is just that the cross-ratio is real $z = \bar{z}$, thus the celestial amplitude must have an overall factor of $\delta(z - \bar{z})$. This is already an indication that, whatever exotic variety of ``conformal field theory'' is to be associated with celestial amplitudes, it must produce interesting new sorts of singularities not usually encountered in familiar local CFTs.\footnote{This arises because of the discrete infinite sequences of primaries implied by translation invariance. While each 2D conformal block contributing to the 4D four-point scattering is a smooth function of $z$, summing over all the primaries produces a delta function \cite{amrs}.}

In recent years, celestial amplitudes  have been intensively studied from a variety of viewpoints \cite{Pasterski:2017ylz,Schreiber:2017jsr,Stieberger:2018edy,Stieberger:2018onx,Fan:2019emx,Pate:2019mfs,Puhm:2019zbl,Nandan:2019jas,Adamo:2019ipt,Guevara:2019ypd,Pate:2019lpp,Law:2019glh,Donnay:2020guq,Albayrak:2020saa,Casali:2020vuy,Casali:2020uvr, Lam:2017ofc}. One theme in these investigations has been the realization that the physics of soft modes in the deep IR -- from the leading and subleading soft theorems to the electromagnetic and gravitational ``memory effects'' -- are beautifully understood from enhancements of global to local conformal and $U(1)$ symmetries on the celestial sphere \cite{He:2014laa,Kapec:2014opa,He:2014cra,Strominger:2014pwa,He:2015zea,Pasterski:2015zua}. However it is also interesting to note that the celestial amplitudes are highly sensitive to UV physics -- there is no Wilsonian decoupling at play. Consider for instance the Mellin transform of the $2 \to 2$ tree-level amplitude in gravity. Since the amplitude grows as the square of the energy, the Mellin transform is simply ill-defined! The scattering of boost eigenstates strongly violates the basic Wilsonian decoupling intuition, since the states we are scattering involve arbitrarily high energy particles. Thus the natural basis for celestial scattering makes it impossible for us to work with only partially consistent, low-energy effective theories: we must deal with fully consistent UV theories from the get-go.

It is thus exciting that an understanding of the physics of the IR, and a proper accounting of UV completion, are both forced on us by the nature of the questions naturally posed on the celestial sphere. This raises the obvious hope that consistency conditions on scattering amplitudes may take a different, more tractable form in celestial terms. Recasting the difficulties associated with the inter-relationship of analyticity and causality for the scattering of momentum eigenstates may open up new routes to directly determining consistent $\cal S$-matrices for general theories and,  especially, for the real world -- to which all the considerations of this paper directly apply!

With this motivation, in this paper we will investigate various basic aspects of celestial amplitudes, from the UV to the IR. 
In section \ref{sec:UV} we study the four-point celestial amplitude $\ca(\beta,z)$, where the ratio of Mandelstam invariants $z=-t/s$ is the 2D conformal cross ratio and 
$\beta =\sum_k(\Delta_k-1) $, in a general effective theory with Wilsonian cutoff $\Lambda_{UV}$. 
In quantum field theory, $\ca$ is shown to be a meromorphic function in the complex $\beta$ plane with poles confined to even integers on the real axis. Poles on the positive axis arise from UV asymptotics, while the residues of the poles on the negative axis are related to coefficients of higher-dimension operators in the low-energy effective action, and are subject to many positivity constraints \cite{Adams:2006sv,deRham:2017avq, Paulos:2017fhb, Guerrieri:2018uew, Correia:2020xtr,Tolley:2020gtv,Guerrieri:2020bto,Caron-Huot:2020cmc,ArkaniNew}.  Multiple poles cleanly reflect the running of higher-dimension operators from massless loops in the low-energy theory. In a complete quantum theory of gravity, one expects high energy  scattering to be dominated by black hole production, which requires exclusive amplitudes to be exponentially damped. This kills all UV divergences, and  all the poles on the positive axis are  accordingly erased! This is  a direct  analog for flat space holography of the erasure of the bulk point singularity in AdS holography \cite{Maldacena:2015iua}, and further underscores the intimate relation between flat space and AdS holography.  For ${\rm Re}( \beta)\to +\infty$, $\ca$ is argued to grow exponentially with a coefficient governed by the black hole area law, while for ${\rm Re}( \beta)\to -\infty$ off the real axis, $\ca$ dies like a power governed by the lowest mass scale in the theory.  Some stringy examples are analyzed in detail. 

In section \ref{sec:IR} we address the IR divergences from photon loops which are regulated in momentum space by a cutoff $\Lambda_{IR}$. 
The well-known exponentiation of these divergences in QED \cite{Weinberg:1965nx} is shown, for massless scalars,\footnote{We hope to report on the technically more challenging massive fermions elsewhere, using the recent construction of a conformal basis for massive fermions \cite{Muck:2020wtx,Narayanan:2020amh}. Collinear divergences from massless scalars are suppressed herein.}  to imply an exact factorization of the celestial amplitudes of the form 
\be \ca=\ca_{\rm soft}\ca_{\rm hard}. \ee
Moreover $\ca_{\rm soft}$ is shown to be the boundary correlator on the celestial sphere of the Goldstone bosons of spontaneously broken large gauge symmetry.  After a shift in the boost weight $\Delta_k$ of the $k$th charge $e_k $ particle proportional to the cusp anomalous dimension
\be \D_k\to \D_k + {e_k^2 \over 4 \pi^2}\ln { \Lambda_{IR}},\ee
it is found that $\ca_{\rm hard}$ is a completely IR-safe quantity. Similar results pertain to gravity, with 
the Goldstone boson for spontaneously  broken supertranslations replacing the one for broken large gauge symmetry in QED.  In section \ref{sec3.3} we consider scattering amplitudes which are rendered IR safe by the inclusion of FK dressings by soft photon clouds \cite{Chung:1965zza,Kulish:1970ut} and their more physical generalizations introduced in \cite{Kapec:2017tkm}. In the celestial basis, conformal symmetry singles out a preferred dressing which is  shown to be equivalent to the insertion of an exponentiated Goldstone boson. We then find that the conformally dressed scattering amplitudes are simply 
\be \ca_{\rm dressed}=\ca_{\rm hard}. \ee
These provide  IR-safe celestial observables for abelian gauge theory or gravity. 

Conventions and technical details are found in the appendix.  As this work was being written up, overlapping results appeared in \cite{Banerjee:2020zlg,Gonzalez:2020tpi}.

\section{The ultraviolet }
\label{sec:UV}

In this section we will begin an exploration of how the ultraviolet physics manifests in  Mellin space,  particularly in  the relationship between the couplings in an effective theory with Wilsonian cutoff $\Lambda_{UV}$ and the analytic structure of the four-point celestial amplitude. 
\subsection{The anti-Wilsonian paradigm}
The relation between celestial and momentum-space amplitudes is especially simple for four-particle scattering. Defining the (scalar) momentum-space amplitude as\footnote{Equation \eqref{spin_gen} in the appendix is the general formula for spinning particles.}
\be {\bf A} = \ms \times \delta^{(4)}\Big(\sum_{i=1}^4p_i\Big), \ee we show in the appendix that\footnote{In general one might have thought it necessary to study $\ca(\D_k)$ as a function of all 4 conformal weights. Interestingly, one of the constraints of translation invariance \cite{Law:2019glh} is that, up to the universal kinematic prefactor $X$,  it can depend only on their sum $\beta+4$. This would not be the case for the 4-point function  in a garden variety  2D CFT.} 
	\be
		\ca (\Delta_i, z_i, \bz_i)  = X {\cal A} (\beta, z),
	\ee
where \be \beta \equiv \sum_{i=1}^4 (\Delta_i - 1) \ee and 
\begin{equation}\label{amp}
{\cal A}(\beta, z) = \int_0^\infty \frac{d \omega }{\omega} \omega^{\beta} \ms(\omega^2, -z \omega^2).
\end{equation}
Here  $X$ is a kinematic factor  fixed by the particle helicities and  derived  in the appendix.  The detailed form of  $X$ is irrelevant  to  our main interest,  namely the $\beta$-dependence in ${\cal A}$. This  is simply given by a single  Mellin transform of the stripped momentum-space amplitude $\ms$ with respect to the center of mass energy, and contains the dynamical information of celestial four-particle amplitudes.

It is easy to see that some basic assumption about UV completeness  is needed in order for the celestial amplitudes to be well-defined and have decent analytic properties. Consider
the celestial amplitude corresponding to the leading 4-point tree amplitude for any theory with a dimensionful coupling constant, where \be \ms (\omega) \propto \omega^{p}.\ee Here   $p \geq 0$  is the number of derivatives in the interaction and $\om$ an overall (fixed angle) energy scale.  The  Mellin transform involves a factor \be 
\ca (\beta )\propto\int_0^\infty \frac{d \omega}{\omega} \omega^{\beta + p}.\ee Clearly $\ca$ is ill-defined, unless we set $\beta = - p + i b$ which gives  $\delta(b)$. But this  is obviously non-analytic in $b$. On the other hand, if the amplitudes are softened to vanish in the deep UV, we get an entirely different celestial amplitude with a nice analytic structure in $\beta$. The fact that the amplitudes are thus  highly sensitive to UV physics is a first example of how celestial amplitudes violate our Wilsonian intuition from momentum-space amplitudes; it is a simple reflection of the fact that we are scattering boost eigenstates that involve arbitrarily high energies.  Since celestial amplitudes involve scattering at all energies, they cannot be fully understood within the traditional Wilsonian paradigm. Nevertheless, we will see both that the powerful manifest symmetries of celestial scattering afford a great deal of control, and that Wilsonian structures are manifest  in a novel way in the analytic behavior of the scattering. 

In the rest of this section we make some preliminary investigations into how the physics of the UV is encoded in the analytic structure of the 4-point celestial amplitude  $\ca(\beta)$ as a function in the complex $\beta$ plane.  There is surely also a yet-uncovered wealth of information in the $z$ dependence of amplitudes, and indeed an investigation of singularities in $z$ is a standard strategy in understanding CFTs.  But it is the transform from $\omega$ to $\beta$ that most directly reflects the novelty of scattering boost eigenstates, and this is the zeroth order aspect of the physics we will focus on herein. As we will see, ${\cal A}(\beta)$ has an infinite tower of poles and interesting characteristic behavior at large $\beta$ which sharply captures  fundamental aspects of the UV physics and reveals  a striking distinction between field-theoretic and quantum-gravitational UV behavior.

\subsection{Poles in $\beta$: field theory vs. quantum gravity}

To begin with, suppose we have a theory of a massless scalar $\phi$ with a cubic coupling $\mu \phi^2 \chi$ to a heavy scalar $\chi$. Integrating out $\chi$ at tree-level gives us an effective quartic coupling $\lambda = - \mu^2/M^2$ together with an infinite tower of irrelevant/higher-dimension operators. The $s$-channel contribution to the stripped four-point tree-amplitude  is \be \ms(p_1,p_2,p_3,p_4) = \lambda \frac{M^2}{s - M^2} . \ee Here 
$s=-(p_1+p_2)^2=\omega^2$ with $\om$ the center-of-mass energy. This  goes to $-\lambda $ at low energies, but dies as $1/s$ at high energies. 

The $\om$-integral \eqref{amp} for the Mellin transform
\begin{equation}
\label{mtpole}
{\cal A}(\beta) = \int_0^\infty \frac{d \omega}{\omega} \omega^\beta  \lambda \frac{M^2}{\omega^2 - M^2}
\end{equation}
can be easily computed by contours, yielding
\begin{equation}\label{cam}
{\cal A}(\beta) =\lambda  M^{\beta}\frac{i\pi}{e^{-i \beta \pi}-1}.
\end{equation}

	This simple example already illustrates the anti-Wilsonian character of the celestial amplitudes.  If we take the momentum-space amplitude and send $M\to \infty$ keeping $\lambda$ fixed, the amplitude
	goes to $-\lambda$, that of the low-energy $\lambda \phi^4$ theory obtained by integrating out $\chi$.  Naively, nothing of the sort happens to ${\cal A}(\beta)$, where dimensional analysis implies that all of the $M$-dependence is in the overall factor $M^\beta$!\footnote{Of course we can formally define the $M \rightarrow \infty$ limit of the Mellin transform by first taking the $M \rightarrow \infty$ limit at the level of the integrand in \eqref{mtpole}. Similarly, for $\beta = i b$, one can regard  
	\be
		\frac{i \pi  }{e^{b\pi}-1}  \lim_{M \to \infty}  M^{i b} =\frac{i \pi }{ e^{b\pi}-1}  \lim_{M \to \infty} \frac{1}{\Gamma (ib)}\int_0^\infty \frac{d \omega}{\omega} \omega^{ib} e^{- \omega/ M}=- 2\pi \delta (b)
	\ee  
	and thereby reproduce the Mellin transform of the $\lambda \phi^4$ amplitude, which was just $-2 \pi \lambda \delta (b)$. Note however, that at any finite $M$, there is a qualitative difference: $\delta(b)$ is non-analytic in $\beta$, with support only at $\beta =0$, while the Mellin amplitude is meromorphic with infinitely many poles away from $\beta =0$. This is in sharp contrast with the amplitude in momentum space, where at fixed $s$, the difference between the full amplitude and that of the low-energy $\lambda \phi^4$ theory vanishes as $M \to \infty$.}  Hence celestial amplitudes are anti-Wilsonian
in that they are highly sensitive to UV physics.  In particular, unlike the case of (low-energy) momentum-space amplitudes,  truncated EFT expansions give drastically different celestial amplitudes, which really 
	only exist in a  formal sense. Once again, this is because scattering boost eigenstates involves arbitrarily high energies, so there is no ``decoupling'' of massive particle states.

Note that the celestial amplitude \eqref{cam} has an infinite tower of poles at $\beta = 2 n$ for all integer $n$. The reason for the presence of these poles and their physical interpretation are extremely simple. Consider a general Mellin transform, and ask for the singularities that can appear from the low-energy $\om \to 0$ part of the integral. To begin with let's assume the amplitude is analytic around $\omega=0$, so we can expand it in powers of $\omega^2$
\begin{equation}
\ms (\omega,z) = \sum_{n=0}^\infty a^{IR}_{n} (z) \omega^{2n}.
\end{equation}
This is of course always true at tree-level. It is also true when we have only integrated out massive particles (that may well only appear in loops), but haven't included the logarithms from massless loops reflecting the running of these coefficients in the low-energy theory.  We will return to dealing with their interesting effects later. Now, let's look at the low-energy contribution to the Mellin transform,
\begin{equation}
\int_0^{\omega_*} \frac{d \omega}{\omega} \omega^{\beta} \sum_{n=0}^\infty a^{IR}_{n} (z) \omega^{2 n}
\end{equation}
where we have cut off the integration at some arbitrary energy $\omega_*$ to emphasize that we are only considering singularities in $\beta$ that arise from the IR as $\omega \to 0$. Obviously, if $a^{IR}_0 \neq 0$,  the integral will have a divergence as
$\beta \to 0$, where the integral as $\omega \to 0$ gives us the pole $a^{IR}_0/\beta$. The integral itself is not convergent for larger negative values of $\beta$, but once we subtract this pole at $\beta=0$ we can analytically continue to more negative values of $\beta$. The new integral converges for $\beta > -2$; there is then a pole at $\beta = -2$ given by $a^{IR}_2/(\beta+2)$, and we can continue in this way to find all the poles in $\beta$. This shows that poles of the analytically continued Mellin amplitude are $\sum_n \frac{a^{IR}_{n}}{\beta+ 2 n}$ and are all simple. Thus the residues of the poles for $\beta = - 2 n$, $n \ge 0$,  are given by the coefficient $a^{IR}_{n}(z)$ of $\omega^{ 2 n}$ in the low-energy expansion of the amplitude ${\bf M}(\omega,z)$.

Similarly, at asymptotically high energies, in any field theory at tree-level, we expect a power-law fall-off for the amplitudes, and so an expansion of the form $\ms(\omega,z) = \sum_m a^{UV}_{m}(z) \omega^{-2m}$. Correspondingly we have poles at $\beta \to + 2m$, whose residues are given by $a^{UV}_{m}$. Hence in field theory  we can have simple poles at all real even integer values of $\beta$, with those on the positive (negative) axis coming from the UV (IR). These general facts are easily verified in our specific example with ${\bf M}(\omega) =    \lambda \frac{M^2}{\omega^2 - M^2}$.

Note however that the high-energy behavior of the amplitude in string theory and quantum gravity is radically different than this field theory power-law fall-off with energy. Already at tree-level, the high-energy amplitudes in string theory scale as $e^{-\alpha^\prime \omega^2}$. More generally, high-energy scattering at energies parametrically larger than the Planck scale is dominated by black hole production. As such, the exclusive $2 \to 2$ scattering amplitude at center of mass energy $\omega$ is expected to be exponentially small, ${\bf M}(\omega, z) \sim e^{-S_{BH}(\omega)/2}$, where $S_{BH}(\omega)= 4\pi G_N \omega^2$ is the entropy of a black hole with mass $\omega$.  A simple model that captures some of the relevant physics for black hole production is the following. We suppose that around mass $\omega$, there are $ \sim e^{S_{BH}(\omega)}$ black hole microstates $|BH_I\rangle$, and that the exclusive amplitude connecting these to the initial state is $\langle BH_I | in\rangle \sim e^{-S_{BH}(\omega)/2} e^{i \phi_I^{in}}$, for some phase $\phi_I^{in}$.  An analogous expression holds for $\langle BH_I | out \rangle$. This tells us that the probability for black hole production is 
\be \sum_{I=1}^{e^{+S_{BH}}} |\langle in |BH_I\rangle|^2 \sim e^{-S_{BH}(\omega)} \times e^{+S_{BH}(\omega)} \sim 1,\ee as expected. The exclusive $2\to2$ amplitude is 
\be 
\sum_I \langle out|BH_I \rangle \langle BH_I |in\rangle \sim e^{-S_{BH}(\omega)} \sum_I e^{i (\phi_I^{in} - \phi_I^{out})}.
\ee
The second factor is a sum over $e^{+S_{BH}}$ phases; we expect the usual $\sqrt{N}$ cancellation in the sum over $N$ random phases, so this sum should have a modulus of order $e^{S_{BH}(\omega)/2}$. This heuristic argument gives us the expected result that
$|{\bf M}(\omega)|^2 \sim e^{-S_{BH}}$. Note that the exponentially small result is a consequence of summing exponentially many phases:  we don't expect the actual $2 \to 2$ amplitude to be a smooth function of energies and angles, but rather to have chaotic fluctuations about this overall exponentially damped envelope.

Hence we expect the high-energy amplitude to fall off as ${\bf M}(\omega) \sim e^{-2\pi G_N \omega^2}$.\footnote{The numerical coefficient in the exponent could be smaller if for some reason for $\om\to \infty$ a finite fraction of the incoming energy does not go into the black hole 
\cite{Giddings:2009gj,East:2012mb}, but the precise coefficient is not important for the following. }  This means that no matter how large and positive the real part of $\beta$ is made, the Mellin integral is well-defined due to this exponential softness in the UV. This leads to a striking qualitative difference between celestial amplitudes in quantum gravity versus UV-complete quantum field theories. In field theory, the celestial amplitude has poles on both the positive and negative real axis, associated with the IR and UV power-series expansions of the amplitude. But in quantum gravity, there are no poles or singularities of any kind in the  Re$(\beta)>0$ half-plane. On the other hand the poles at negative $\beta$, reflecting the expansion of the low-energy amplitude, are present for both field theory and gravity.

We can see the difference between power-law and exponentially soft  high-energy behavior explicitly in a toy example, where we imagine ${\bf M}(\omega) \propto e^{-\omega^2/M^2}$. This simple example is unphysical, running afoul of causality/unitarity positivity constraints, but it suffices to illustrate the difference in pole structure.  The Mellin transform gives the $\Gamma$ function,
\begin{equation}
\int_0^\infty \frac {d \omega}{\omega} \omega^{\beta} e^{- \omega^2/M^2} = \frac{1}{2}M^\beta \Gamma(\beta/2),
\end{equation}
which indeed only has poles on the negative real axis at $\beta = - 2 n$, for $n \geq 0$, and is otherwise analytic everywhere in the right half-plane. The residues on these poles, ${\rm Res}\left[ \frac{1}{2}M^\beta \Gamma(\beta/2)\right]_{\beta = - 2 n} = M^{-2n} e^{in\pi}/n!  $, are exactly the coefficients of the $\om$ power-series expansion of ${\bf M} \propto e^{-\omega^2/M^2} = \sum \om^{2n}M^{-2n} e^{in\pi}/ n!  $, as expected from our general discussion above.

The absence of poles on the positive $\beta$ axis is a flat space analog for celestial amplitudes of the absence of the ``bulk point'' singularity for AdS/CFT correlators \cite{Maldacena:2015iua}. The discussion of \cite{Maldacena:2015iua} was given in terms of position-space boundary correlators, but the analogy with what we have seen for celestial amplitudes is cleaner to see in the boundary momentum-space version of their discussion. Consider a boundary correlation function $F(\vec{k}_i)$. It is easy to see that Witten diagrams for massless particles in $AdS$ have a singularity when the sum of the ``energies'', $E_{tot} = \sum_i |\vec{k}_i|$, is analytically continued, with some of the $|\vec{k_i}|$ negative and some positive, allowing $E_{tot} \to 0$. The vectors $\vec{k}_i$ now specify the kinematics of a scattering process, and the coefficient of the singularity as $E_{tot} \to 0$ can be interpreted as the high-energy limit of the flat-space scattering amplitude \cite{Raju:2012zr,Maldacena:2011nz}. Schematically, the singularity arises from bulk integrals where all the interaction vertices go to infinity together, giving us dependence on $E_{tot}$ as $\int^\infty dz e^{-E_{tot} z} A_{flat}(z \vec{k}_i)$ where $A_{flat}(z \vec{k}_i)$ approaches the flat-space scattering amplitude with blue-shifted momenta $z \vec{k}_i$ as $z \to \infty$. Now if we just have field theory in the bulk, $A_{flat}(z \vec{k}_i)$ falls off at most as a power at large $z$.  Thus the integral over $z$ must have a singularity at $E_{tot} \to 0$, since for $E_{tot}<0$ the exp$(-E_{tot} z)$ factor dominates and the integral necessarily diverges. But, if the bulk theory has gravity as we have discussed, the high-energy scattering amplitude is dominated by black-hole production and falls off as exp$(-c z^p)$ with $p=\frac{(D-2)}{(D-3)} >1$ in $D>2$ spacetime dimensions. 
Hence the integral is completely analytic in $E_{tot}$, since exp$(-E_{tot} z)$ is dominated by exp$(-c z^p)$ and the integral is convergent for any $E_{tot}$, so there is no singularity as $E_{tot} \to 0$. This is exactly the same mechanism that removes all poles for positive $\beta$ in celestial amplitudes.

\subsection{Logarithmic running and higher order poles in $\beta$}

Let us now consider logarithms which arise in the quantum theory. As we have already discussed, the amplitudes have logarithmic IR divergences which require an IR  cutoff $\Lambda_{IR}$. For the real-world  scattering of photons and gravitons,  these divergences naturally exponentiate, and can be universally stripped off. In the next section we will study these IR divergences and show they take a very simple form in Mellin space. In an effective theory with  a Wilsonian cutoff $\Lambda_{UV}$,  loops of photons and gravitons also result in logarithmic UV divergences which we study in this section.  The low-energy expansion in powers of $\omega^{2n}$ is further modulated by powers of $(G_N \omega^2){\rm  log} (\omega^2/\Lambda_{UV}^2)$. For instance, if we consider photon-photon scattering, at 1-loop from photon-graviton loops we find for the $(1^-2^+3^-4^+)$ helicity amplitude
\begin{equation}
	\begin{split}
{ \ms}^{-+-+} &= \frac{G_N^2  \langle 1 3 \rangle^2 
	 [2 4]^2 }{240}\left[\left (-7 + 290 \left({s \over u}\right) +90 \left({s \over u}\right)^2 + 60\left ({s \over u}\right)^3\right) \log \left(\frac{s}{\Lambda_{UV}^2}\right)  \right. \\ 
	&\quad \quad - \left. \left(267 + 290\left ({s \over u}\right) + 90\left ({s \over u}\right)^2 + 60\left ({s \over u}\right)^3 \right) \log\left(\frac{t}{\Lambda_{UV}^2}\right)  +{\rm UV \, finite} \, \right],
	\end{split}
\end{equation} 
where $t=-zs$ and $u=(z-1)s$. While the detailed coefficients in this expression are not important, the structure of this amplitude nicely illustrates some well-known but crucial aspects of effective field theory. The expression contains logarithms, 
$\log (s)$ and $\log (t)$.  The coefficients of these logarithms are calculable from the IR physics and unaffected by the UV uncertainties associated with the precise nature of the cutoff $\Lambda_{UV}$. This is because the discontinuities across the logarithmic branch cuts are determined by unitarity through cutting the diagram, given in this case by products of tree amplitudes for photons and gravitons. The coefficient of $\log(\Lambda_{UV}^2)$ is easily seen to simply be $ (137/120) G_N^2 \langle 13 \rangle ^2[24]^2$.  Note that all the dependence on the interesting ratios $s/u=1/(z-1)$ drops out, as must physically be the case: this divergence is entirely equivalent to the amplitude we would get from  a local $F^4$ contact term, which indeed has the trivial kinematical dependence $\langle 13 \rangle^2[24]^2$. We can think of this UV logarithm as giving a ``running'' of the $F^4$ operator, logarithmically induced by the gravitational coupling $G_N$.

More generally,  in the theory of photons and gravitons, the general form for the low-energy expansion of the scattering amplitude has a logarithmic modulation with powers of the effective gravitational coupling $(G_N\omega^2)$ and the large logarithms log($\omega/\Lambda_{UV})$:
\begin{equation}\label{weff}
\ms =   \sum_{r \leq m; n}  a_{n,m,r}(z) \omega^{2 n} \times (G_N \omega^2)^m {\rm log}^r (\omega/\Lambda_{UV}),
\end{equation}
where $m$ is the number of loops. 
We know that the leading term $n=1,m=r=0$ in this expansion is just controlled by the low-energy theory of photons plus gravity. As we have seen, the leading large logarithms, with $r=m$, are also captured by photon-graviton loops, with an intricate set of in principle-calculable predictions for the $z$-dependence of the coefficients. The coefficient of the $n=2,m=r=0$ term for photon-photon scattering (the contact $F^4$ term) does depend on the UV theory. Note that while the simple contact $F^4$ term already   gives the trivial dependence on $\omega,z$,  the finite part of the 1-loop amplitude gives a calculable but more intricate dependence on $z$. Also a contact $F^4$ term, dressed by photon/graviton loops, will induce subleading logarithmic corrections. This  expansion contains a wealth of physical information, some of which is calculable in the low-energy theory, but much of which carries information about the UV completion.

This familiar organization of how we think about scattering amplitudes in terms of ``the analytic part'' reflecting contact interactions and the ``non-analytic parts with logarithms'' is even more sharply reflected in Mellin space.
If we look again at the possible divergences in the celestial amplitude coming from the $\omega \to 0$ region of integration, we have only to note that
\begin{equation}
\int_0^{\omega_*} \frac{d \omega}{\omega} \omega^a {\rm log}^b \omega \sim  \frac{\partial^b}{\partial a^b} \int_0^{\omega_*}  \frac{d \omega}{\omega} \omega^a \sim \frac{1}{a^{1 +b}}.
\end{equation}
Thus, the presence of the logarithms in the low-energy theory is reflected by higher-order poles at $\beta = -2n$. Indeed, we have that
\begin{equation}
{\cal A}(\beta \to -2N,z) \to  \sum_{n=1}^N \sum_{r=0}^{{N-n}} \frac{1}{(\beta + 2N)^{r+1}}  a_{n,m=N-n,r}(z).
\end{equation}
Note that we only have a simple pole at $\beta = -2$, up to a double pole at $\beta = -4$ and poles up to $N$th order at $\beta=-2N$. The full structure of the low-energy expansion of the scattering amplitude, including both the analytic and non-analytic dependence on energy, is beautifully captured by this specific (multiple)-pole structure in Mellin space.

\subsection{Positivity  constraints}

This connection allows us to make a start at formulating some of the known consistency conditions on amplitudes, following from unitarity and causality, in Mellin space. For momentum-space amplitudes, causality implies the existence of dispersive representations for amplitudes at fixed $t$; this gives a partial wave expansion at fixed $s$, and unitarity imposes that the coefficients of the Legendre polynomials in this expansion should be positive. Now, let us consider the expansion of the low-energy amplitude, for simplicity here ignoring the massless loops/running in the low-energy effective theory, so we have only ${ \ms}(s,t) = \sum a_{p,q} s^p t^q$. It has long been known that causality and unitarity imply that the coefficients $a_{p \geq 2,q=0}$\footnote{Or in the previous notation $a_{n \geq 2}(z=0)\geq 0$.} should all be positive (see \cite{Adams:2006sv} and references therein). More recently \cite{ArkaniNew}, it has been
realized that this entire table of coefficients $a_{p,q}$ must satisfy a vast number of hidden positivity conditions associated with a ``positive geometry'' known as the EFT-hedron. Constraints sharpen some standard intuitions into precise bounds on the coefficients of ``garden-variety'' higher-dimension operators contributing to $a b \to a b$ scattering amplitudes.  For instance we shouldn't expect  operators of the same mass dimension to have vastly different coefficients. One might think this is merely a consequence of some type of ``naturalness'', and that by suitable fine-tuning in the high-energy theory, we can engineer any relative sizes we like between these operators. However the EFT-hedron shows that this is not the case and not everything goes, forcing the $a_{p,q}$ to satisfy linear inequalities for fixed mass dimension $p+q$. We also expect operators to be suppressed by a similar scale, $i.e.$ not to have dimension 6 operators suppressed by the TeV scale while dimension 8 operators  suppressed by the Planck scale. The EFT-hedron sharpens this expectation as well, imposing non-linear inequalities between different $a_{p,q}$. For instance for fixed $q$, we form the ``Hankel Matrix'' $A_{ij} = a_{i+j,q}$; the claim is that all the minors of the matrix $A$ must be positive. Now we can trivially translate between the $a_{p,q}$ and the $a_{n,m=0,r=0}(z)$ of our low-energy expansion:
\begin{equation}
a_{n,m=0,r=0} (z) = \sum_{p,q;p+q=n}a_{p,q}  (t/s)^q .
\end{equation}
So all positivity constraints on the $a_{p,q}$, directly translate into positivity constraints on the residues of ${\cal A}(\beta,z)$ on its simple poles at $\beta = -2n$. Note that already the most trivial constraint that the residues be positive in the forward limit rules out the simplest guess for a celestial amplitude with poles only for negative $\beta = -2 n$: $\Gamma(\beta/2)$ has this property and was discussed in a toy example as following from the transform of $\exp(-\omega^2)$, but its residues  $(-1)^n/n!$ at $\beta=-2 n$ are alternating in sign. String amplitudes do scale as $e^{-\alpha^\prime \omega^2}$ at fixed angle, but deviate from this in the Regge limit, scaling as $s ~(s^2)$ for open (closed) strings for large $s$ and fixed $t$, consistent with causality constraints.  This is part of the magic of string theory, which manages to give exponentially soft amplitudes for fixed angle scattering, while miraculously satisfying the many positivity constraints associated with causality and unitarity. 

Clearly the positivity constraints on the EFT-hedron translate into powerful and hard-to-satisfy constraints on the 
analytic structure of the full quantum celestial amplitude $\ca(\beta, z)$.  It would be interesting to translate all the non-linear positivity constraints on the EFT-hedron into constraints on the pole and asymptotic behavior of the meromorphic function $\ca(\beta,z)$.  We leave a more complete characterization of these constraints
and their solutions to future work. 

\subsection{Large $\beta$ behavior}

It is also interesting to consider the behavior of the celestial amplitude at large $\beta$. The behaviors as Re$(\beta) \to \pm \infty$ are  associated with different physics, and we consider them in turn.

Let's begin with $\beta \to + \infty$, and assume that  the high-energy growth of scattering is eventually dominated by black-hole production. For large $\om$, we expect the modulus $|{\bf M}(\om)|\sim e^{- S_{BH}(\om)/2}$, and the amplitude ${\bf M}(\omega)$ itself to be modulated by a randomly oscillating phase, so that ${\bf M} \sim e^{-S_{BH}(\omega)/2} e^{i \phi(\omega)}$. Then the Mellin integral is  convergent for all ${\rm Re} (\beta) > 0$ and dominated by large $\omega$ as  $\beta \to + \infty$. The leading behavior for $|{\cal A}(\beta)|^2$ is
\begin{equation}
|{\cal A}({\beta \to \infty})|^2 \sim \int^\infty \frac{d \omega}{\omega} \omega^{\beta} e^{-2 \pi G_N \omega^2}
\int^{\infty}\frac{d \omega^\prime}{\omega^\prime} \omega^{\prime \beta} e^{-2 \pi G_N \omega^{\prime 2}} e^{i (\phi(\omega) - \phi(\omega^\prime))}.
\end{equation}
The phase $\phi(\omega)$ is expected to oscillate wildly and the oscillations wash out the integral, localizing it to a one-dimensional integral.  Thus at large $\beta$, we have 
\begin{equation}
|{\cal A}(\beta \to \infty)|^2 \sim \int^\infty \frac{d \omega }{\omega^2} \omega^{2 \beta} e^{-4\pi G_N \omega^2} \sim G_N^{-\beta} \Gamma(\beta).
\end{equation} 
Thus the modulus of the celestial amplitude grows factorially with $\beta$. It is interesting that while the $2 \to 2$ amplitude is exponentially {\it small} at high energies, the corresponding celestial amplitude is instead (slightly more than) exponentially {\it large} at large positive $\beta$! 

Let's now switch to discussing the behavior as $\beta \to - \infty$. Note that unlike the exercise we just performed above, we can't trivially analyze the limit starting with the Mellin integral, since the integral doesn't converge there and must be defined by analytic continuation. Nonetheless,  as we will see, the large $\beta$ behavior is very simple to characterize. To get a quick idea of the physics involved, let's return to our simple first example of an amplitude obtained by integrating out a massive scalar $\chi$ with a cubic coupling to $\phi$, generalized to  several $\chi_i$'s with different masses $M_i$. The celestial amplitude is then
\begin{equation}
{\cal A}(\beta) = \sum_i \lambda_i\frac{M_i^\beta i \pi }{  e^{-i\pi  \beta}-1}.
\end{equation}
Now, let's go to large $\beta \to - \infty$. Clearly, it is the {\it lightest} of all the states, with mass $M_{lightest} \equiv M_*$,  that will dominate the celestial amplitude in this limit, and we have the behavior
\begin{equation}
{\cal A}(\beta \to - \infty) \sim M_{*}^{\beta} \times \frac{1}{e^{-i\pi  \beta}-1}.
\end{equation}
Let us now try to understand this result more directly, avoiding explicit computation. The factor $1/(e^{-i\pi  \beta}-1)$ must be there in any celestial amplitude, as it accounts for the infinite tower of poles for $\beta = - 2n$.
The factor $M_{*}^\beta$ is also easy to understand. Let's consider the low-energy expansion of the stripped amplitude as $\sum_n a_n \omega^{2n}$. We know that this expansion has a finite radius of convergence, and breaks down when $\omega^2 \to M_*^2$. Furthermore, we know that when $\omega^2$ is very close to $M_*^2$, it is well approximated by the pole $1/(\omega^2 - M_*^2) = -1/M_*^2 \sum_n (\omega^2/M_*^2)^n$. Thus, while we can't say anything universal about the coefficient $a_n$ when $n$ is small, we learn that at large $n$, we must have that $a_n \sim \frac{1}{M_*^{2n}}$.

Now recall that the $a_n$ are also the residues of the celestial amplitude at $\beta = - 2 n$. At large negative $\beta$, the residue of the celestial amplitude at $\beta = - 2n$ must then scale as $M_*^{-2n} = M_*^\beta$, and thus $M_*^\beta/(e^{-i\pi  \beta}-1)$ has the correct poles and residues for large $\beta$ and correctly captures the asymptotic behavior of ${\cal A}(\beta)$.

This logic holds more generally. Suppose instead of a tree-level UV completion, we integrated out massive particles at loop level.  It will be slightly more convenient to work with  $s = \omega^2$. The amplitude $\ms(s)$ can have branch points either for positive $s$ (corresponding to the threshold for $s$-channel particle production) or negative $s$ ($u$-channel production) but for simplicity let's only consider $s$-channel production; as will become obvious in a moment in general whichever branch point is closest to the origin will dominate. Suppose the $s$-channel threshold starts at $s=M^2_*$ . Now, as usual we can express the coefficients $a_n$ as a contour integral around the origin $a_n \sim \oint \frac{ds}{s^{n+1}} \ms(s)$, and then deform the contour around the branch cuts in the $s$ plane (in this field-theoretic example for large $n$ there is no contribution from the contour at infinity). Then $a_n$  is expressed as
\begin{equation}
a_n \sim \int_{s=M_*^2}^\infty \frac{ds}{s^{n+1}} \times {\rm Disc}(\ms)(s).
\end{equation}
Since the factor $1/s^{n+1}$ is  rapidly dying for large $n$, this integral becomes dominated near threshold. The near-threshold discontinuity across the cut, or what is the same, the near-threshold total cross-section, will be well-approximated as ${\rm Disc}(\ms) \sim (s - M_*^2)^q$ for some power $q$ depending on the precise coupling to the massive particles.  Inserting this into the integral lets us conclude that for large $n$, $a_n \to M_*^{-2n} n^{-(1 + q)}$. This has a simple analytic dependence on $n$, and so we conclude that asymptotically for large and negative $\beta$,
\begin{equation}
{\cal A}(\beta) \to M_*^\beta \beta^{-(1 + q)} \times \frac{1}{e^{-i\pi  \beta}-1}
\end{equation}
has all the correct poles and residues at large $\beta$. Thus the leading behavior is again given by $M_*^\beta$, with the subleading correction $\beta^{-(1 + q)}$ reflecting the power-law growth of the production cross-section of new states near threshold.

We expect similar behavior at large negative $\beta$ in string theory. First, we describe how to apply the previous analysis of the large negative $\beta$ behavior of tree-level amplitudes to tree-level string amplitudes. Famously, the fixed-angle amplitude  ${\bf M}(s,z)$ cannot be expressed as an infinite sum over $s$ poles with bounded residues; the pole expansion can be done only by keeping one of $s,t,u$ fixed. Nonetheless, the amplitude ${\bf M}(s)$ for fixed $z=-t/s$ of course does have an infinite number of poles in $s$. The closest pole to the origin is at $s=M_s^2$ where $M_s$ is the mass of the lightest string state, and the dependence of the amplitude on $s$ approaches $\frac{r(z)}{s - M_s^2}$. Thus if we look at the low-energy expansion of $\ms(s,z)= \sum_n a_n(z) s^n$, we must again find that $a_n \to r(z) M_s^{-2n}$ in order to correctly capture the breakdown of this expansion at $s=M_s^2$.
So the leading large $\beta$ behavior of the corresponding celestial amplitude $\ca(\beta,z)$ will again scale as $M_s^\beta \times \frac{1}{e^{-i\pi  \beta}-1}$.

We can easily confirm this asymptotic behavior for the coefficients $a_n$ explicitly from the string amplitude.  To do so, we use 
an argument based on dispersion relations similar to the one before. Just for simplicity we will consider open-string scattering, and consider the scattering of massless colored scalars so as to ignore the irrelevant details of spin. The scattering amplitude is
\begin{equation}
{\ms} = s^2 \times \frac{\Gamma(- s) \Gamma(-t)}{\Gamma(1 -s - t)} .
\end{equation}
Working at fixed real $z=-t/s$ with $0<z<1$ in the physical region,  we have $\ms(s,z) = s^2 \Gamma(-s) \Gamma(s z)/\Gamma(1-s(1 - z))$. Note that the residues of $\ms(s,z)$ as $s \to m$ decrease exponentially at large $m$, while the residues of the poles at $t \to m \implies s \to -m/z$ grow exponentially at large $m$, which is why we can't express $\ms(s, z)$ as a sum over its poles. Here $m$ are non-negative integers, $m = 1,2,3 \cdots$. However it is still easy to obtain the coefficients in low-energy expansion $\ms(s,z)= \sum a_n(z) s^n$, just as above. We again express the coefficients $a_n$ as the usual contour integral $a_n(z) = \frac{1}{2 \pi i } \oint \frac{ds}{s^{n+1}} \ms(s,z)$. Now we deform this contour to one that runs parallel to the imaginary axis, to the left of the origin and the right of the first pole at negative $s$, then running counterclockwise at infinity for large positive real part of $s$. Note that $\ms(s,z)$ decreases as we go to infinity along the vertical line in both directions, and vanishes on the circle at infinity. So, when further suppressed by the factor of $1/s^{n+1}$, this contour integral vanishes at large $n$. We are left with the sum over poles at $s = m$ ($m = 1, 2, \cdots$), where the residues decrease with $n$ as $\sim m^{-n}$, and so at large $n$, we are dominated by the residue at $s=1$. 

We can also confirm the asymptotic large $\beta$ behavior by direct inspection of the Mellin transform of the string amplitude.    To do this it is  useful to use the Euler integral for the Beta function. However, this integral is only well-defined when $s,t$ are both negative, so we must analytically continue to get to the physical region. For the purposes of discussing the large $\beta$ behavior of the Mellin transform this detail is not important; we will therefore directly Mellin transform with respect to negative $s$. A related analytic continuation was performed in \cite{Stieberger:2018edy}. We now set $t/s=z$ with $z>0$, and study
\begin{equation}
\hat{{\cal A}}(\beta, z) = \frac{1}{2} \frac{e^{i\pi\frac{ \beta}{2}}}{1 + z}\int_0^\infty \frac{ds}{s} s^{\frac{\beta}{2} + 1} \int_0^1 \frac{dy}{y(1-y)} y^s (1-y)^{z s},
\end{equation}
where the ``hat'' is to remind us that this is the Mellin transform with respect to negative $s$. Writing $y^s (1 - y)^{z s} = \exp (- s[\log 1/y + z \log 1/(1-y)])$, we can
perform the integral first over $s$ and arrive at
\begin{equation}
\begin{split}
\hat{\mathcal{A}}(\beta, z) &= \frac{1}{2}\frac{e^{i\pi\frac{ \beta}{2}}}{1 + z}\Gamma \left(\frac{\beta}{2} + 1\right)  \int_0^1 \frac{dy}{y(1-y)} \left( \log \left(\frac{1}{y}\right) + z \log \left(\frac{1}{1-y}\right) \right)^{-(\frac{\beta}{2} + 1)} \\
&\equiv \frac{1}{2}\frac{e^{i\pi\frac{ \beta}{2}}}{1 + z} \Gamma\left(\frac{\beta}{2} + 1\right) \times I.
\end{split}
\end{equation}
Note that the integral $I$ converges for $\beta > 0$, but requires an analytic continuation for the large negative $\beta$ we are interested in. However it is easy to analytically continue the integral $I$.

The integral diverges from the behavior near $y \to 0$ (giving us the $s$-channel poles of the amplitude) and $y \to 1$ (giving the $t$-channel poles). Let $y_*$ denote the intermediate point on the interval $(0,1)$ where $\log(1/y_*) = z \log(1/(1-y_*))$. We can then divide the integral $I = I_< + I_>$ into the regions between $(0,y_*)$ and $(y_*,1)$, and appropriately Taylor expand as: 
\begin{equation}
I_< = \int_0^{y_*} \frac{dy}{y} \left(\log\frac{1}{y} \right)^{-(\frac{\beta}{2} + 1)} \times (1 + y + y^2 + \cdots) \times \left[1 + \frac{z(y + \frac{y^2}{2} + \cdots)}{\log\frac{1}{y}} \right]^{-(\frac{\beta}{2} + 1)}
\end{equation}
and
\begin{equation}
\begin{split}
I_> = \int_{y_*}^{1} \frac{dy}{1-y} \Big(z \log\frac{1}{1-y}\Big)^{-(\frac{\beta}{2} + 1)} &\times (1 + (1-y) + (1-y)^2 + \cdots) \\
&\times \left[1 + \frac{(1-y) + \frac{(1-y)^2}{2} + \cdots}{z \log\frac{1}{1-y}}\right]^{-(\frac{\beta}{2} + 1)}.
\end{split}
\end{equation}
Note that only the leading term in the expansion of $I_<$,  $I^0_< = \int_0^{y_*} \frac{dy}{y} (\log\frac{1}{y})^{-(\frac{\beta}{2} + 1)}$ is divergent for $\beta \leq 0$. Similarly only the leading term in the expansion of $I_>$, $I^0_> = \int_{y_*}^1 \frac{dy}{1 - y} (z \log\frac{1}{1-y})^{-(\frac{\beta}{2} + 1)}$ is divergent for $\beta \leq 0$. We can integrate to find
\begin{equation}
I^0_> + I^0_< = \frac{2 + 2z}{z} \frac{\left(\log\frac{1}{y_*}\right)^{- \beta/2}}{\beta}
\end{equation}
which can be analytically continued past the pole at $\beta = 0$ to all $\beta$. Note that this grows exponentially for large negative $\beta$ but, as we will now see, is
subleading to the remaining part of $I$.

The remaining terms in the expansion in $y$ given above are already analytic in $\beta$. At large negative $\beta$, the dominant contribution is
\begin{equation} \label{leadingI}
I \to I_<-I_<^0 \to  \int_0^{y_*} dy \left(\log\frac{1}{y}\right)^{-\frac{\beta}{2}  - 1} \to \Gamma \left(- \frac{\beta}{2}\right)
\end{equation}
leading to
\begin{equation}
\hat{{\cal A}}(\beta) =\frac{1}{2} \frac{e^{i\pi\frac{ \beta}{2}}}{1 + z} \Gamma\left(\frac{\beta}{2} +1\right) \times I \to \frac{1}{2} e^{i\pi\frac{ \beta}{2}} \Gamma\left(\frac{\beta}{2}+1\right) \times \Gamma \left( - \frac{\beta}{2}\right) = \frac{i \pi }{e^{-i\pi \beta} - 1} .
\end{equation}
Thus we recover the expected poles in $\beta$, with no other $\beta$ dependence, precisely reflecting the expected behavior that for $\ms(\omega) = \sum_n a_n \omega^{2n}$, we have $a_n \to 1$ as $n \to \infty$.  Note that in our analysis we were working with units with $M_s \to 1$.  Restoring this would give us the expected overall $M_s^\beta$ dependence. 
In \eqref{leadingI}, we found that only $I_<$ contributes to the leading behavior of $I$.  The leading large negative behavior of $I_>$ is associated
with the $t$-channel poles, to which a similar argument applies.   Here  
\be
	I_> -I_>^0\to \int_{y_*}^1 dy \left(z\log\frac{1}{1-y_*}\right)^{-\frac{\beta}{2}  - 1} \to z^{-\frac{\beta}{2}}\Gamma \left(- \frac{\beta}{2}\right),
\ee
so we find an extra suppression by a power of $z^{-\frac{\beta}{2}}$ when $z <1$.  This  just reflects the fact that the large $\beta$ asymptotics is controlled by the proximity of poles to the origin; working at fixed $z= t/s$, the poles as $t \to m$  are at $s = m/z $ and thus the first $t$-channel pole is farther away from the origin by this factor $z^{-1}$.

\subsection{What determines ${\cal A}(\beta,z)?$}

We have understood some of the basic analytic properties of ${\cal A}(\beta,z)$. The (multiple) poles and residues on these poles have a nice interpretation associated with the low-energy effective field theory expansion, which must satisfy positivity properties due to causality and unitarity. The behavior at large positive $\beta$ scales as ${\cal A}(\beta \to + \infty) \sim \Gamma(\beta/2)$ and reflects black hole dominance of the high-energy $\cal S$-matrix, while the leading behavior ${\cal A}(\beta \to - \infty) \sim M_*^{\beta}/(e^{-i\pi \beta}  - 1)$ is controlled by the lightest threshold for new physics. It is likely that there is more information contained not just in the leading behavior as $\beta \to -\infty$, but in the structure of all the subleading corrections to this leading behavior as well. After all even at tree-level, the leading behavior as $\beta \to -\infty$ is controlled only by the lightest of the states in the UV, and presumably the presence of all the states is reflected as  some expansion in exponentials of the form $M_i^{\beta}$.

Can we determine ${\cal A}(z,\beta)$ just by specifying the residues on all its poles, together with its asymptotic behavior? Strictly as a mathematical fact, the answer is clearly ``no''. For any ${\cal A}(z,\beta)$ having the correct poles and asymptotics, we can always add e.g. ${\cal A}(z,\beta) \to {\cal A}(z,\beta)+ c \mu^{\beta}$. This doesn't affect any of the poles and so long as $\mu>M_*$, this correction is subdominant to the leading behavior for ${\cal A}(z,\beta)$ as $\beta \to \pm \infty$. However, it is interesting to note that literally this trivial shift cannot possibly correspond to sensible physics. If we transform back to momentum space, it corresponds to $\ms\to \ms + 2 c \delta(s - \mu^2)$. This is brutally non-analytic in $s$, and so violates the most basic tests of causality.

Thus the fundamental challenge confronting any ab-initio theory for the $\cal S$-matrix -- understanding how even a rough notion of locality and causality for the scattering process is encoded in an observable measured at infinity -- must also be dealt with for celestial amplitudes. There must be additional constraints on ${\cal A}(\beta,z)$ that encode causality, amongst other things guaranteeing decent analytic structure back in momentum space.  We hope that the simple observations in this section on the analytic structure in the $\beta$ plane will serve as a jumping off point for further explorations of this fascinating question.

\section{The infrared}	
\label{sec:IR}

Our discussion so far has skirted around the ubiquitous IR divergences arising in QED, non-abelian gauge theory and gravity which require the introduction of an IR cutoff $\Lambda_{IR}$. As $\Lambda_{IR} \to 0$, the amplitude for any finite number of particles goes to zero, reflecting the dominant emission of infinitely many soft quanta. For momentum-space amplitudes, there is a well-known exponentation of soft divergences, and it is possible to ``strip-off'' these soft IR divergences. 

In this section, we will see that this exponentiation of IR divergences has a beautiful celestial avatar for QED and gravity.\footnote{It is amusing to remark, again, on a qualitative difference between momentum-space amplitudes and celestial amplitudes. For undressed momentum-space  amplitudes, some type of ``stripping'' of IR divergences is necesssary, since the amplitudes simply vanish as $\Lambda_{IR} \to 0$. But this is not the case for the celestial amplitude! Consider the case of gravitational four-point scattering discussed in section \ref{sec:3.2}; thanks to the exponentiation we know that ${\cal A}(\beta) = \int_0^\infty \frac{d \omega}{\omega} \omega^\beta {\rm exp}[- |\gamma| c(z) G_N \omega^2] {\az}(\omega)$. But this expression does not simply vanish as the $\Lambda_{IR} \to 0, \gamma \to -\infty$. We can see this in the toy example of the Gamma function, where we consider $\int_0^\infty \frac{d \omega}{\omega} \omega^\beta {\rm exp}(-|\gamma| \omega^2) = |\gamma|^{-\beta/2} \Gamma(\beta/2)$. This vanishes as $\gamma \to -\infty$ for $\beta>0$, but diverges for negative $\beta$. Furthermore, if we look at the poles on the negative $\beta$ axis of celestial amplitudes,  they are deformed by powers of $|\gamma| c(z) G_N$, with a fixed $z$ dependence. Thus there is IR cutoff independent content even in the ``unstripped'' celestial amplitude, unlike for momentum-space amplitudes.} 
In the first  subsection we address these in detail in (massless scalar) QED, and demonstrate the simple hard/soft factorization in a conformal basis.  The second subsection treats the gravitational case. The third   subsection  considers IR finite  Fadeev-Kulish (FK) dressed amplitudes in both QED and gravity, along with their more physical generalizations which do not require incoming dressings,  introduced in \cite{Kapec:2017tkm}. The conformal primary condition is shown to single out a unique dressing from the infinite family of FK dressings. Moreover it can be compactly represented by an insertion of the Goldstone boson for the spontaneous breaking of large gauge or supertranslation symmetry.  A simple and universal formula is presented for IR safe celestial amplitudes. 	

\subsection {Scalar QED}

In this subsection we  consider massless scalar QED, as a model  for studying the IR divergences  associated to photons in 
real-world QED with electrons. We accordingly ignore   IR divergences from the  massless scalar loops. We expect similar conclusions pertain to the physical but technically more involved case of massive fermions, whose  conformal primary wavefunctions have been constructed only recently  \cite{Muck:2020wtx,Narayanan:2020amh}. We hope to report elsewhere on the massive case.

The most general quantum theory of massless QED is described in momentum space by   a Wilsonian effective action with a cutoff  at a scale $\Lambda_{UV}$ which we take to be lower than any other scale in the theory. The effective interactions of photons and scalars with momenta $p_i$ below the cutoff $\Lambda_{UV}$  and charges  $eQ_i$\footnote{Here $Q_i$ are integers obeying $\sum_k Q_k=0$.} are characterized by a double  expansion in $\om/\Lambda_{UV}$ and $\ln (\om/\Lambda_{UV})$.  Physical amplitudes are then corrected by loops of photons with momenta below $\Lambda_{UV}$, and do not depend on $\Lambda_{UV}$.\footnote{They do of course depend on physical high-energy mass scales in the theory.}  Here  we encounter IR divergences from exchanges of soft photons between charged external lines and must introduce an IR regulator $\Lambda_{IR}$ on the internal loop momenta. These divergences are known to exponentiate \cite{Weinberg:1965nx} and the momentum-space  scattering amplitude takes the form
\be \label{qsa}
{\bf A} = e^{B } {\az},
\ee
where ${\az}(p_i)$  does not involve $\Lambda_{IR}$ or $\Lambda_{UV}$. The exponential prefactor is\footnote{Here we will not address collinear divergences arising from massless charges but
drop their contribution to $B$ in a Lorentz-invariant manner.  Including contributions from collinear divergences, $B$ is of the form
\be
	B = - \frac{1}{2} \alpha \sum_{i,j} Q_i Q_j \ln [p_i\cdot p_j/\mu^2] = - \frac{1}{2} \alpha \sum_{i,j} Q_i Q_j \ln (-2 \eta_i \eta_j \omega_i \omega_j z_{ij} \bz_{ij} /\mu^2)   ,
\ee where $\mu$ is an arbitrary mass scale.  Invoking total charge conservation $\sum_k Q_k = 0$,   the above can be put in the form
\be
	B = - \frac{1}{2} \alpha \sum_{i,j} Q_i Q_j \ln ( \omega_i \omega_j z_{ij} \bz_{ij}  )   =  - \frac{1}{2} \alpha \sum_{i,j} Q_i Q_j \ln \left| \frac{1}{2} p_i \cdot p_j \right| ,
\ee
which upon dropping the collinear-divergent terms (i.e. when $i =j$), is precisely the expression in \eqref{qexp}.  While the $\omega_i$'s in the logarithm could also have been removed by this argument, 
we choose to keep them so that  the result upon discarding the collinear-divergent terms  is Lorentz invariant.   We will see that  the  $\omega_i$'s give rise to a shift in conformal dimensions which is required by ${\rm SL} (2, \mathbb{C})$ covariance. 
}	\be\label{qexp}
		\begin{split}
			e^B &= e^{- \alpha \sum_{i< j} Q_i Q_j \ln |\frac{1}{2}p_i \cdot p_j|},
				\end{split}
	\ee
	where 
	\be \alpha = \frac{e^2}{4 \pi^2} \ln  \Lambda_{IR} \ee
is the cusp anomalous dimension times the logarithm of the IR cutoff. 	
	Taking the Mellin transform  we find
	\be\label{rfg}
		\begin{split}
		 \ca (\Delta_i, z_i, \bz_i)&=
			\prod_{j } \int_0^\infty d \omega_j ~ \omega_j^{\Delta_j -1}  e^{- \alpha \sum_{k< l} Q_k Q_l \ln | \frac{1}{2}p_k \cdot p_l|}{\az}\\
			& =\Big(\prod_{m<n}  ( z_{mn} \bz_{mn} )^{-\alpha Q_mQ_n}\Big)\prod_j\int_0^\infty d \omega_j~ \omega_j^{\Delta_j -1}
				\Big(\prod_{k<l} (\omega_k \omega_l)^{-\alpha Q_k Q_l} \Big) {\az}.\\
		\end{split}
	\ee
	Here we have used $p _i \cdot p_j=-2 \eta_i \eta_j \omega_i \omega_jz_{ij} \bz_{ij} $, with $\eta_i=\pm 1$ for  outgoing  (incoming) energy $\omega_i$ particles which cross the celestial sphere at $z_i$, and $z_{ij}=z_i-z_j$.
		
		Next we use $\sum_iQ_i=0$ to write 
	\be
	\prod_{i<j} ( \omega_i \omega_j)^{-\alpha Q_i Q_j}
				= \prod_{i} \omega_i^{\alpha Q_i^2}.\ee
This enables us to rewrite \eqref{rfg} in the conformally soft factorized form 
	\be \label{sfc}
	 \ca = \ca_{\rm soft} \ca_{\rm hard},\ee
where the  hard\footnote{Note that the  hard factor involves a Mellin transform of external states whose energies are taken all the way to zero, and are not cut off at $\Lambda_{IR}$ even though internal loop momenta are. For such an object there are no loop corrections to  the subleading soft graviton theorem \cite{Cachazo:2014dia} which implies 2D conformal invariance.}
and soft factors are 
	\be  \ca_{\rm soft}\equiv \prod_{i<j} ( z_{ij} \bz_{ij} )^{-\alpha Q_i Q_j},\ee
	\be \ca_{\rm hard}\equiv \prod_{i }\int_0^\infty d \omega_i ~ \omega_i^{\Delta'_i -1}  {\az },~~~~~~\Delta'_i\equiv \Delta_i + \alpha Q_i^2.\ee 
 In terms of the renormalized  conformal dimensions $\Delta'_i$, the `hard' amplitude $ \ca_{\rm hard}$ is exactly what it would have been had we simply omitted the IR-divergent factor $e^B$ in \eqref{qsa}. So the effect of including the soft photon loops is simply a shift  renormalization of  the $\D_i$ by  the cusp anomalous  dimension $\alpha Q_i^2$ and a multiplication of  the amplitude by the IR-divergent factors $ \ca_{\rm soft}$.  

We see  that celestial amplitudes conformally factorize and
 IR-safe information about  the physical observables is encoded in $\ca_{\rm hard}$.  IR-divergent factors directly cancel out in certain  ratios of amplitudes such as those that give the electromagnetic memory effect. Moreover in subsection \ref{sec3.3.3} we will see that $\ca_{\rm hard}$ are in fact nothing but the conformal-primary-dressed celestial Faddeev-Kulish amplitudes.

This conformally soft amplitude $ \ca_{\rm soft}$  can be written as a 2D correlator by introducing a  2D free boson $\Phi\sim \Phi+2\pi $ with a two point function proportional to the cusp anomalous dimension
\be\label{rtd} \< \Phi(z, \bz)\Phi(w,\bw)\>= \alpha\ln(z-w)+ \alpha\ln(\bz-\bw).\ee
One finds for an $n$-point amplitude 
\be \label{ssx} \ca_{\rm soft}=\< e^{iQ_1\Phi(z_1, \bz_1)}  \cdots e^{iQ_n\Phi(z_n, \bz_n)}\>   .  \ee
The free boson $\Phi$ was introduced already in \cite{Nande:2017dba}. It is the Goldstone boson
for spontaneously broken large gauge symmetry and will be further discussed below.

\eqref{sfc} expresses conformally soft factorization of celestial amplitudes. It is well known that, with a carefully defined separation of scales, momentum space amplitudes in QED factorize into a product of hard and soft parts. Simple factorization does not automatically  apply in an arbitrary  basis -- sums of momentum eigenstates do not in general factorize. It is not a priori obvious that any such factorization need occur for conformal basis celestial
amplitudes. Indeed \eqref{sfc} is not the usual hard-soft energy factorization - energy is not even an argument of the $\cs$-matrix! Rather it is a conformally hard-soft factorization, whose implications for the construction of an IR-safe  $\cs$-matrix will be discussed in section \ref{sec3.3}.

	\subsection{Gravity}
	\label{sec:3.2}
	In this section we consider $2\to2$ scattering of gravitons in a general quantum theory of gravity with assumed soft high-energy behavior.    As for QED, we describe this by a Wilsonian effective action with a cutoff at a scale $\Lambda_{UV}$ which is lower than any other scale in the theory. The effective action is  characterized by a double  expansion in $\om/\Lambda_{UV}$ and $\ln (\om/\Lambda_{UV})$. Amplitudes are then corrected by loops of gravitons with momenta below $\Lambda_{UV}$.  Here  we encounter IR divergences from exchanges of soft gravitons between external lines and must introduce an IR regulator $\Lambda_{IR}$ on the internal loop momenta. These divergences are known to exponentiate and, as in the QED case \eqref{qsa}, the scattering takes the form \cite{Weinberg:1965nx}
\be \label{qsag}
{\bf A} = e^{B } {\az},
\ee
where $\az$ does not involve $\Lambda_{IR}$ or $\Lambda_{UV}$.
However instead of \eqref{qexp} the exponent is \cite{Naculich:2011ry}
\be \label{stx}
\begin{split}
B &= -\g \sum_{i, j} (p_i \cdot p_j) \ln (p_i\cdot p_j)\\  
&= 2\g \sum_{i, j} \eta_i \eta_j \omega_i \omega_j |z_{ij}|^2 \ln |z_{ij}|^2,
\end{split}
\ee
with
\be \g= {G \over \pi}\ln \Lambda_{IR} .\ee
To get the second line in \eqref{stx} we used momentum conservation
$
\sum_i p_i = 0$ and 
\be p_i \cdot p_j=-2 \eta_i \eta_j \omega_i \omega_j |z_{ij}|^2 .\ee 
The celestial amplitude is the Mellin transform of \eqref{qsag}
\be \label{caf}
\ca=\prod_{k } \int_0^{\infty} d\omega_k\omega_k^{\Delta_k- 1} \exp{\left[2\g \sum_{i,j} \eta_i
\eta_j \omega_i \omega_j |z_{ij}|^2\log|z_{ij}|^2  \right]}{\az}.
\ee
Denoting by $G(p_k)$ the graviton operator with momentum $p_k$, we may define  the translation  operator $P_k$ which acts  as
\be P_kG(p_k)= \eta_k\om_k G(p_k).\ee
Acting on conformal primaries this becomes
\be P_kG_{\D_k}(z_k,\bz_k)=\eta_kG_{\D_k+1}(z_k,\bz_k),\ee
while  on a celestial amplitude
\be P_k\ca(\D_1,z_1,\bz_1;...\D_n,z_n,\bz_n)=\eta_k\ca(\D_1,z_1,\bz_1;...\D_k+1,z_k,\bz_k;...\D_n,z_n,\bz_n).\ee
Using this operator we may pull the exponential out of the integrand in \eqref{caf} and write
\be \ca=\ca_{\rm soft}\ca_{\rm hard}.\ee
Here the hard amplitude is exactly the un-soft-dressed expression
\be
\ca_{\rm hard}=\prod_{i } \int_0^{\infty} d\omega_i~ \omega_i^{\Delta_i - 1}{\az},
\ee
and is an IR safe quantity.
The conformally soft amplitude $\ca_{\rm soft}$ is an operator
\be \ca_{\rm soft}=\exp{\left[2\g \sum_{i,j}  P_i P_j |z_{ij}|^2\log|z_{ij}|^2  \right]},\ee
which shifts the conformal weights of the scattering states in the hard amplitude.
$\ca_{\rm soft}$ contains all the IR divergences.

The fact that the soft amplitude is an operator, rather than just a number, is familiar even in momentum space. Only for the leading soft photon factor in momentum space is the soft factor simply a number. The subleading soft photon factor involves the angular momentum, whose representation in a momentum basis is an operator which  differentiates the amplitude with respect to the momentum. Here we see that the leading soft graviton factor is a number in a momentum basis, but not in a conformal basis where it shifts the boost weight.

As in the gauge theory case, $\ca_{\rm soft}$ can be represented as a correlator of Goldstone bosons.
The Goldstone boson for supertranslations is the metric component at the boundary of ${\cal I}$ conventionally denoted $C(z,\bz)$. Its celestial two point function was computed in \cite{Himwich:2020rro} and shown to be determined from the gravitational cusp anomalous dimension as 
\be \<C(z,\bz)C(w,\bw)\>= -4\g |z-w|^2\ln|z-w|^2.\ee
While charged operators in QED have the dressings $e^{iQ_k\Phi(z_k,\bz_k)}$, energetic operators in gravity in a momentum basis are dressed by $e^{i\eta_k \om_kC(z_k,\bz_k)}$. In a conformal basis this becomes  $e^{iP_kC(z_k,\bz_k)}$, and we have
\be\label{asg} \ca_{\rm soft}=\<e^{iP_1C(z_1,\bz_1)}\cdots e^{iP_nC(z_n,\bz_n)}\>,\ee
in analogy to the gauge theory case \eqref{ssx}. Note however that in the gravity case IR safe quantities are obtained simply by stripping off $\ca_{\rm soft}$,  without a simultaneous  renormalization of the conformal dimensions.

\subsection{IR-safe dressed amplitudes}
\label{sec3.3}

 In this section, we explain how to obtain infrared finite\footnote{Up to suppressed collinear and loop divergences associated with massless charged particles, ignored throughout this paper.} celestial amplitudes by dressing charged states in a conformal basis with photon or graviton clouds of the Goldstone modes constructed in \cite{Donnay:2018neh}.  The dressings studied herein
are a special conformally invariant choice of Faddeev-Kulish (FK) dressing \cite{Chung:1965zza,Kulish:1970ut} and render amplitudes infrared finite by the usual FK analysis, with one important difference: the soft IR-divergence-cancelling clouds are centered at the same point on the celestial sphere as the associated hard particle, but the cloud can be either ingoing or outgoing independently of whether the hard particle is ingoing or outgoing. This important generalization does not spoil the cancellation of IR divergences  \cite{Kapec:2017tkm,Choi:2017ylo} and is required for  the inclusion of physically relevant incoming configurations with hard but no soft particles. 

We begin by identifying a conformally invariant representative of  the family of Faddeev-Kulish dressings in momentum space.
Then, we show that the $\Delta=1$ Goldstone mode from \cite{Donnay:2018neh} is the \emph{only} definite-conformal-weight  mode of the photon that appears with this choice of dressing.
Finally, we explain, for hard massless particles,  how the  dressing transforms when the hard charges are put in a conformal basis.  

\subsubsection{Faddeev-Kulish dressings}

Exclusive scattering amplitudes of charged particles with finite numbers of external photons vanish because they violate the conservation laws following from  large gauge symmetries \cite{Kapec:2017tkm,Choi:2017ylo}.  At the level of Feynman diagrams, this  is implemented by IR divergences which set the amplitudes to zero.
Faddeev and Kulish  and others  \cite{Chung:1965zza,Kulish:1970ut} constructed a set of dressed states with non-vanishing, IR-safe scattering amplitudes. The FK states are comprised of  charged particles dressed by soft photon clouds which restore the conservation laws by setting all large gauge charges (except the global one\footnote{ For brevity we restrict the following discussion to the incoming states with zero net global charge. A very similar treatment applies to the more general case. }) to zero. Large gauge charge is not carried by any finite number of photons so these states necessarily contain a divergent number of photons near $\om=0$. Shortcomings  of the FK states, which have hindered their general usage, are that they both do not span the Hilbert space of physical states and are overcomplete on the subspace they do span. These shortcomings are discussed and remedied below.  The  large family of IR-safe FK dressings of the (massless or massive) single particle state $ |p_k,Q_k\>$ of charge $eQ_k$ and momentum $p_{k\mu}$ are written
\be W_k[f ]|p_k,Q_k\>,\ee
where
\be\label{wkf}		\begin{split}
			W_k[f ] =\exp \left[ -  e Q_k
						\int \frac{d^3 \vec q}{ (2 \pi)^3} \frac{f(\vec q)}{2 q^0}
							 \left( \frac{p_k \cdot   \ve^{*\alpha}  }{  p_k \cdot q } a_\alpha (\vec{q}) -\frac{p_k \cdot   \ve^{\alpha} }{p_k \cdot q} a^\dagger_\alpha (\vec{q})   \right)\right]
		\end{split}
	\ee
	and $f(\vec q)$ is any function which obeys
\be f(0)=1.\ee
We use here the standard plane-wave mode decomposition for the field operator $\hat{A}$
\be\label{amode}
		\hat{A}_\mu (x) = e \int \frac{d^3 \vec q}{ (2 \pi)^3} \frac{1}{2 q^0} \left[\ve^{*\alpha}_\mu  a_\alpha (\vec{q}) e^{i q \cdot x} + \ve^{\alpha}_\mu  a^\dagger_\alpha (\vec{q}) e^{-i q \cdot x} \right],
	\ee
	with conventions given in the appendix.
Parameterizing both $q$ and the polarization vectors $\ve^\pm$ by $(\om, z,\bz)$ as in  \eqref{4mompar} and \eqref{polpar}, the dressing may be written  \be \label{wilson_f}
		\begin{split}
			W_k[f]&=\exp \left[ - \frac{e Q_k  }{ \sqrt{2}(2 \pi)^3}    \int_0^\infty d \omega\int d^2z  f(\om,z,\bz)
							 \left( \p_\bz \ln (p_k \cdot  \hat q) \left(  a_+ (\omega, z, \bz) -  a^\dagger_-(\omega, z, \bz) \right)  \right.\right. \\& \quad \quad \quad
							 \quad \quad \quad \quad \quad  \quad \quad \quad  \quad \quad \quad  \quad  \quad
							 \left.	\left.+\p_z \ln (p_k \cdot  \hat q)  \left(  a_- (\omega, z, \bz) -  a^\dagger_+(\omega, z, \bz) \right) \right)  \right].
		\end{split}
	\ee
	
\subsubsection{Goldstone dressing}	
Consider  the special choice of dressing 
	\be\label{dsf} f(\vec q)=1,\ee
	for which the $\om$ integral in \eqref{wilson_f} reduces to  $\D=1$ Mellin transforms of $a_{\pm}$ and $a^\dagger_{\pm}$. 
	In this subsection we show that 
	$W_k[1]$  then becomes  an exponential of the Goldstone mode $\Phi$ discussed earlier. FK did not consider $f(\vec q)=1$ because finiteness of energy requires $f\to 0$ for $|\vec q|\to \infty.$ Such a constraint  is not relevant in a conformal basis. 
	
We first review the conformal Goldstone wavefunction and its canonical pairing with the soft mode \cite{Donnay:2018neh}.
The conserved Klein-Gordon norm for
spin-1 massless fields 
	 \be
	 	(A, A') = -i \int_{\Sigma} d\Sigma^\mu \left [A^\nu F'^*_{\mu\nu}- A'^*{}^\nu F_{\mu\nu}\right]
	 \ee
	is often easily evaluated when the Cauchy slice $\Sigma$ is taken to be $\ci^+$ or $\ci^-$.  This norm pairs ingoing or outgoing positive-helicity conformal primary wavefunctions of weight $(h,\bar h)=(1+{i\l\over 2}, {i\l\over 2})$ for $\l $ real but nonzero with themselves.\footnote{A similar discussion applies to the negative-helicity weight $(h,\bar h)=({i\l\over 2}, 1+{i\l\over 2})$ which will be suppressed here for brevity.} Several special things happen at $\l=0$  \cite{Donnay:2018neh}.\footnote{We omit in the following discussion yet another pair of currents associated to large magnetic gauge symmetry \cite{Strominger:2015bla, Nande:2017dba} which complexify $J_z$ and $S_z$.} The weight $(1,0)$ wavefunction obtained in the limit $\l\to 0$, is  denoted $A_{\mu}^{\rm G}(x;w,\bw)$ and referred to as the Goldstone wave function. It is  real, has vanishing norm, and is pure gauge.  Moreover the same mode is obtained from the limit of either ingoing or outgoing modes. Since there is only one mode, at first the canonical partner seems to be missing. However explicit analysis of the Maxwell equation for $(h, \bar h)=(1,0)$  reveals the appearance of a second  `conformally soft' solution, denoted $A_{\mu}^{\rm CS}(w,\bw)$
which is the canonical partner of $A_\mu^{\rm G}(z,\bz)$:
\be\label{csg}(A^{\rm CS}(z,\bz),A^{\rm G}(w,\bw) )=\frac{2\pi i}{e^2}\delta^{(2)}(z-w).\ee
 $A^{\rm CS}$ can be constructed as a limiting difference of modes near $\l=0$ and is not pure gauge. It describes a certain radiative shock wave moving along the light cone of the origin \cite{Donnay:2018neh}.
The explicit wavefunctions are\footnote{These conformally soft wavefunctions defined here are solutions to Maxwell's equations and differ from the ones in \cite{Donnay:2018neh} by a logarithmic term which doesn't affect our analysis and is dropped here.}
\be
		A_{\mu}^{\rm G} (x; w, \bw) =\frac{1}{e^2} \p_\mu \left [- \frac{x \cdot \p_{\bw} \hat q(w, \bw)  }{x \cdot \hat q(w, \bw) }\right],
	\ee
\be	
		\begin{split}
			A_{\mu}^{\rm CS}(x;w,\bw)  & = -\frac{e^2}{4\pi}\Theta (x^2) A_{\mu}^{\rm G}(x;w, \bw).
		\end{split}
	\ee
	
The mode of the field operator $\hat{A}$
 \be	  \label{gs1}
	 	\begin{split}
			J_{z}& = \left (\hat{A}, A^{\rm G}(z,\bz)\right)
		\end{split}
	 \ee
	 is the conformally soft photon current.\footnote{Note that because  the pairing \eqref{csg} is off-diagonal,  $J_z$ is multiplied by the wavefunction $A^{\rm CS}$ in the expansion of the field operator $\hat{A}$.} Its Ward identities are the conformally soft photon theorem.  $J_z$ generates, but is invariant under, large gauge transformations.
A second current, which shifts under large gauge transformations and is  referred to as the Goldstone current, is defined by
 \be	  
	 	\begin{split}
			S_{z}& = \left (\hat{A}, A^{\rm CS}(z,\bz)\right) .
		\end{split}
	 \ee
 Defining the plane wave solutions
\be A^{\pm, {\rm P}}_{\mu;w} (x; \omega, w, \bw)  =  \ve^{+}_\mu (w, \bw)  e^{\pm i \omega \hat q (w, \bw) \cdot x} ,\ee
\be A^{\pm, {\rm P}}_{\mu;\bw} (x; \omega, w, \bw)  = \ve^{-}_\mu (w, \bw)  e^{\pm i \omega \hat q (w, \bw) \cdot x} ,\ee
one has
\be
		\begin{split}
			\left (A_{\bw}^{\pm , {\rm P}}, A^{\rm CS}(z,\bz) \right)  &= \mp \frac{1}{\sqrt{2}\omega} (2\pi) \delta^{(2)} (z-w),
			\\
			\left (A_w^{\pm , {\rm P}}, A^{\rm CS} (z,\bz)\right)  &= \pm \frac{1 }{\sqrt{2}\omega (z-w)^2}.
		\end{split}
	\ee
Using these relations the Goldstone current is expressed in plane wave creation and annihilation operators as 
	 \be	
	 	\begin{split}
			S_{z}= -\frac{e}{\sqrt{2}(2\pi)^2} \int_0^\infty d \omega   \Big[   a_+(\omega, z, \bz)&-  a^\dagger_-(\omega, z, \bz) \\
			& - \int\frac{ d^2 w}{2 \pi } \frac{1}{(z-w)^2} \left( a_- (\omega, w, \bw)  -  a^\dagger_+(\omega, w, \bw)  \right)   \Big] .
		\end{split}
	 \ee
It is straightforward to show that
			$\p_z S_\bz = \p_\bz S_z$ and $S_{\bz}^\dagger=-S_z$. It follows that the hermitian
			Goldstone boson $\Phi$ defined by
			\be S_z=i\p_z\Phi\ee
has  the explicit expression
			\be
			\begin{split}
		\Phi (z, \bz) = \frac{ie}{\sqrt{2}(2\pi)^2} \int_0^\infty d \omega \int \frac{d^2w}{2 \pi }
					  &\Big[  \frac{1}{\bz-\bw} \left (a_+ (\omega,w, \bw)-   a^\dagger_- (\omega, w, \bw) \right)\\
				 	&~~~~~\qquad+ \frac{1}{z-w} \left( a_- (\omega,w, \bw)  -  a^\dagger_+ (\omega, w, \bw)  \right)   \Big].
				 	\end{split}
	\ee
	 Then, for the special conformal  choice $f(\vec q)=1$, after integration-by-parts \eqref{wilson_f} becomes
	 \be  
	 \label{dressing1}
		\begin{split}
			W_k [1]&=\exp \left[ \frac{e Q_k  }{ \sqrt{2}(2 \pi)^3  }    \int d^2z \ln \left[p_k \cdot \hat q (z,  \bz) \right] \right. \\&   \quad \quad \quad \quad  \left.
							\times\int_0^\infty d \omega \left[   \p_\bz \left (  a_+ (\omega, z, \bz) -  a^\dagger_-(\omega, z, \bz) \right) +\p_z \left (  a_- (\omega, z, \bz) -  a^\dagger_+(\omega, z, \bz) \right)  \right]  \right]\\
			& = \exp \left[ -  iQ_k       \int \frac{d^2z}{2 \pi}  \ln \left[p_k \cdot \hat q (z,  \bz) \right]  \p_z \p_\bz \Phi \right]\\
			& =  \exp \left[ -   i Q_k  \int  d^2 z  ~G_{2} ( p_k; z, \bz) \Phi (z, \bz)  \right],
		\end{split}\ee
 		where $G_2$ is the bulk-to-boundary propagator on the hyperbola $H_3$ given in \cite{Nande:2017dba}.
\eqref{dressing1} is the Hermitian conjugate of the Goldstone boson dressing of momentum-space massive charged particles discussed in \cite{Nande:2017dba}. 
Hence we have shown that conformal FK dressings are Goldstone boson insertions. 

 	In the ultrarelativistic (or massless)  limit, $ p_k$ approaches a boundary point $(z_k,\bz_k)$, $G_2\to
	 \delta^{(2)}(z-z_k)$ and
	\be \label{rds} W_k
			\to e^{-   i Q_k  \Phi (z_k, \bz_k) }.\ee
This is the hermitian conjugate of the operator appearing in \eqref{ssx}. The conformal weight of this primary operator is IR divergent and governed by the cusp anomalous dimension as follows from \eqref{rtd}	 and discussed in \cite{Nande:2017dba}. 	One finds  $(h_k,\bar h_k)=(-{\alpha Q_k^2\over 2}, -{\alpha Q_k^2\over 2})$.		

We close this subsection with several comments.
This exact choice of dressing \eqref{dsf} was already used as an example in the discussion of IR divergences in \cite{Kapec:2017tkm}. Since it is conformal primary, it does not have finite energy and would not have been considered by Faddeev and Kulish. However the conformal primary condition neatly eliminates the unwanted ambiguity in the choice of the dressing function
$f(\vec q)$ which inevitably appears if a finite energy condition is imposed.

Finally we note that $W_k[1]$ can alternately be  expressed as a Wilson line of the form
\be \label{wilson1}
		W = \exp \left[- i \int d^4 x ~ \hat{A}_\mu j^\mu\right],
	\ee
	where the source is a charged particle of momentum $p_k$ coming from the origin
	\be
		j^\mu (x) = -Q_k \int_0^\infty d \tau ~e^{ -\eps \tau }p_k^\mu \delta^{(4)} (x^\nu- p^\nu_k \tau),
	\ee
where $\eps\to 0$ is a regulator. Inserting the mode expansion \eqref{amode} and doing the integrals, 	 \eqref{wilson1} becomes
	\be
		\begin{split}
			W_k =\exp \left[  - e Q_k \int \frac{d^3 \vec q}{ (2 \pi)^3} \frac{1}{2 q^0}
							 \
						\left( \frac{p_k \cdot   \ve^{*\alpha}  }{  p_k \cdot q } a_\alpha (\vec{q}) -\frac{p_k \cdot   \ve^{\alpha} }{p_k \cdot q } a^\dagger_\alpha (\vec{q})   \right)\right],
		\end{split}
	\ee
which  agrees with \eqref{wkf} for $f(\vec q)=1$. The relation between Wilson lines and FK dressings was discussed in
reference \cite{Choi:2018oel}.

	\subsubsection{IR-finite celestial amplitudes}
	\label{sec3.3.3}
	
		In the previous subsection, we showed that a specific unique choice of  FK dressing for a charged particle in a momentum eigenstate could be identified with dressing by the  Goldstone mode $\Phi$ for large gauge transformations. 
		
		To complete the story,  we now adapt these dressings  for charged particles in a conformal rather than momentum basis, and show that the complete dressed state is a conformal primary. We  consider here only the simpler case of massless charges, hoping to return elsewhere to the more physical massive case.   $p_k$ is then null and the dressing localizes to a point $(z_k,\bz_k)$ on the celestial sphere as in  \eqref{rds}.  The transformation from momentum space to a conformal basis involves only the energy $\om_k$ of the particle, and does not touch the dressing. Hence the dressing trivially still factorizes in a conformal basis. Infrared-finite celestial amplitudes between massless charged particles are obtained by dressing
		the representative operators $\co_{\Delta_k} (z_k, \bz_k)$ with \eqref{rds}.
		However we must address the question of normal ordering. The OPE ($i.e.$ collinear singularity) of the original operator $\co$ and its dressing is,
according to \eqref{rtd} and \eqref{ssx} \cite{Nande:2017dba} (dropping the $k$ subscript)
\be :e^{-iQ\Phi(z,\bz)}::\co_\D(w,\bw):\sim |z-w|^{2\alpha Q^2}:e^{-iQ\Phi(w,\bw)}\co_\D(w,\bw):.\ee
We therefore define the dressed operator $\hat \co$ by
\be \hat \co_{\D+\alpha Q^2}(w,\bw)=\lim_{z\to w} |z-w|^{-2\alpha Q^2}:e^{-iQ\Phi(z,\bz)}::\co_\D(w,\bw):.\ee
$\hat \co$ has no OPEs with the Goldstone boson $\Phi$ or $J_z$. This factorization of correlation functions into a current algebra and ``parafermionic'' piece is familiar in 2D CFT.
IR finite celestial amplitudes are then simply
\be \label{IRfinite_ca}
			 \ca_{\rm dressed} = \langle \hat \co_{\Delta_1} (z_1, \bz_1) \cdots \hat \co_{\Delta_n} (z_n, \bz_n)  \rangle .
		\ee	
The dressed amplitudes differ from the undressed ones by extra terms arising from the Goldstone boson current algebra. Because the exponents in \eqref{ssx} and \eqref{rds} differ by a sign, these exactly cancel the factor of $\ca_{\rm soft}$. Hence we conclude
\be\label{ash}  \ca_{\rm dressed}= \ca_{\rm hard} .\ee

\subsubsection{Completeness}

The original basis of states discussed by Faddeev and Kulish is {\it overcomplete} because states with different dressings $f(\vec q)$ are not orthogonal.  One cannot make the dressed state a momentum eigenstate and there is no canonical choice of $f(\vec q)$. This makes it hard for example to study unitarity. We have seen that in a conformal basis, in contrast, there is a canonical choice of dressed state which is a conformal primary. Restricting to such states eliminates the awkward overcompleteness of the FK states.

A second problem with FK states is that they are {\it incomplete}. All the large gauge charges, except for the one global charge, vanish for every FK state. This is more or less the  defining property of the FK state. In the classical limit, this means that the leading ($ 1 \over r^2$) radial component of the electric field is constant at infinity (i.e. ${\cal I}^+_-$). This condition is violated for a generic configuration of incoming charges: for example an incoming  colliding $e^+e^-$ pair with no incoming photons. Of course, given any incoming set of charges, one can add a radiative electromagnetic field with arbitrarily small energy  fine-tuned to make the leading field constant at infinity. But this is a highly nonlocal, fine-tuned procedure and does not correspond to any  realistic physical situation such as in $e^+e^-$ collisions. Certainly a complete basis should include  states corresponding to incoming charged particles with no incoming radiation. These cannot be FK states. 

Another way to say it is that the restriction to FK states does not properly cluster:  two very widely separated non-FK states, each with non-constant leading radial electric field at large distances, can combine to make an FK state.

A more  complete and physically realistic set of IR-safe scattering amplitudes which  are not fine-tuned or nonlocal  were constructed in
\cite{Kapec:2017tkm,Choi:2017ylo}. When an undressed charged particle  scatters off of its incoming trajectory,
 it produces a beam of  photons which emerge at the point on the sphere at ${\cal I}^+$ antipodal to the one from which the particle emerged in the far past. Large gauge charge conservation - {\it i.e.} the soft photon theorem - requires the outgoing beam to have the same soft poles that would appear in an FK dressing of the incoming particle. In other words, if the FK dressing is not added by hand to the incoming charge, it automatically appears as a photon beam in the outgoing state. Scattering amplitudes that are consistent with this requirement are IR finite and those that are not vanish \cite{Kapec:2017tkm,Choi:2017ylo}. There is no need to impose the unphysical requirement that all charged particles are dressed or equivalently that all large gauge charges vanish.

IR finiteness can be achieved by putting the required  cloud of FK photons in either the in or the out state. In celestial amplitudes, we find the $f=1$ dressing is an exponential of the Goldstone mode which can be equivalently described as either a limit of an ingoing or an outgoing cloud in the absence of other soft insertions. Hence, for a conformal primary cloud, the scattering amplitudes are the same whether it is taken to be incoming or outgoing:  both are given by $\ca_{\rm hard}$.

Our discussion so far omits an important generalization: we are also interested in scattering for which some of the external particles are soft such as insertions of $J_z$.  In that case the soft part of the amplitude involves more than \eqref{ssx} and there are corrections to \eqref{ash} which we have not worked out.  $J_z$ insertions  are needed for example to describe the memory effect and distinguish between in and outgoing clouds of Goldstone modes. (In the undressed formulae, with an explicit IR cutoff, these are given by an IR-safe ratio of amplitudes with and without insertions of $J_z$ \cite{Strominger:2014pwa}.)
We leave this generalization and  a complete prescription for an IR-finite and unitary $\cal S$-matrix to future work. 
\subsubsection{Gravity}
In this subsection we sketch the straightforward (if more complicated) extension of these results to gravity.
More details of identities used in the following appear in \cite{Himwich:2020rro} where the soft $\cal S$-matrix for gravity was rewritten in celestial form.

FK dressed states in gravity take the general form
\be
\label{fk}
|\vec{p}_k\rangle_{\rm dressed} = e^{R_k[f]}|\vec{p}_k\rangle,
\ee
where
\be
\begin{split}
R_k[f] &= -\frac{\kappa}{2}\int \frac{d^3 \vec{q}}{(2\pi)^3} \frac{f(\vec q)}{2 q^0} \frac{p_{k}^{\mu} p_{k}^{\nu}}{p_k \cdot q}\left( \varepsilon^{*\alpha}_{\mu\nu} a_{\alpha}(\vec{q}) - \varepsilon^{\alpha}_{\mu\nu} a^{\dagger}_{\alpha}(\vec{q}) \right)\\ &= \frac{\kappa \eta_k \omega_k}{2 (2\pi)^3} \int d\omega d^2z f(\om,z,\bz)\\
&\qquad \qquad \times \Big( \frac{(z - z_k)}{\bz - \bz_k}(a_+(\omega, z, \bz) - a_-^{\dagger}(\omega, z, \bz)) + \frac{(\bz - \bz_k)}{z - z_k}(a_-(\omega, z, \bz) - a_+^{\dagger}(\omega, z, \bz)) \Big).
\end{split}
\ee
Here $\kappa=\sqrt{32\pi G}$, $f(0)=1$ and the mode decomposition of the field operator is 
\be\label{hmode}
		\hat{h}_{\mu\nu} (x) = \frac{\kappa}{2} \int \frac{d^3 \vec q}{ (2 \pi)^3} \frac{1}{2 q^0} \left[\ve^{*\alpha}_{\mu\nu} (\vec q)  a_\alpha (\vec q) e^{i q \cdot x} + \ve^{\alpha}_{\mu\nu} (\vec q)  a^\dagger_\alpha (\vec q) e^{-i q \cdot x} \right].
	\ee
	Expressions for the polarization tensors $\ve^\pm_{\mu\nu}$ are given in the appendix \eqref{tpar}.
The pure gauge weight $(-{1 \over 2},{3 \over 2})$ Goldstone wavefunctions are \cite{Donnay:2018neh}
\be
h^{\rm G}_{\mu\nu} = \lim_{\Delta \rightarrow 1} h_{\mu\nu}^{\Delta} = \p_{\mu}\xi_{\nu} + \p_{\nu}\xi_{\mu}, \quad \xi_{\mu} \equiv \frac{1}{\kappa^2} \p_{\bz}^2[\hat{q}_{\mu}\log(-\hat{q} \cdot x)].
\ee
The $(-{1 \over 2},{3 \over 2})$ conformally soft modes are
\be
h_{\mu\nu}^{ \rm CS}(x; w, \bw) =\frac{\kappa^2}{8\pi} h_{\mu\nu}^{\rm G}(x; w, \bw) \Theta(x^2),
\ee
and describe  radiative shock waves along the light cone of the origin $x^2=0$. One finds these are canonically paired 
\be
(h^{\rm CS}(w, \bw), h^{ \rm G}(w',\bw'))= -\frac{8\pi i}{\kappa^2} \delta^{(2)}(w - w')
\ee
with respect to the conserved inner product $(~,~)$ for $h_{\mu\nu}$ given in \cite{Donnay:2018neh}.
In analogy with the gauge theory case the supertranslation current is constructed from 
$\left(\hat{h}, h^{\rm G}(z,\bz)\right)$, while  the Goldstone mode $C$ is defined by 
\be
\begin{split}
 i\p_w^2C(w,\bw)&=(\hat{h},h^{\rm CS})\cr
&=\frac{\kappa}{2(2\pi)^2} \int_0^{\infty} d\omega\Big(a^\dagger_-(\omega, w, \bw) - a_+(\omega, w, \bw) \\
&\qquad \qquad \qquad + \int \frac{d^2z}{\pi} \frac{\bz - \bw}{(z - w)^3}\left(a_+^\dagger(\omega, z, \bz) - a_-(\omega, z, \bz) \right) \Big).
\end{split}
\ee
After some algebra we find that, acting on a particle  state of momentum $p_k$,
\be\label{dsm} e^{R_k[1]} =e^{-i\eta_k \omega_kC(z_k,\bz_k)}.\ee
This shows that the special case of FK dressing $f=1$ is, as for QED, equivalent to the  exponentiated Goldstone boson, but with the prefactor of the charge $Q_k$ traded for the energy $\omega_k$. After transforming to the conformal basis, the dressing becomes  an insertion of the operator $e^{-i P_kC(z_k,\bz_k)}$ appearing in $\ca_{\rm soft}$ in equation \eqref{asg}. One then finds, as for the QED case, the amplitudes
\be \ca_{\rm dressed}=\ca_{\rm hard} \ee
 again define IR-safe scattering amplitudes. 

\section*{Acknowledgements}	
We benefited from conversations with Alex Atanasov, Alfredo Guevara, Mina Himwich, Dan Kapec and Walker Melton. This work was supported by DOE grant de-sc/0007870 and by Gordon and Betty Moore Foundation and John Templeton Foundation grants via the Black Hole Initiative.  NAH is supported by the DOE under grant de-sc/0009988. MP acknowledges the support of a Junior Fellowship at the Harvard Society of Fellows. Research at Perimeter Institute is supported in part by the Government of Canada through the Department of Innovation, Science and Industry Canada and by the Province of Ontario through the Ministry of Colleges and Universities.

\appendix
\section{Conventions}

	Null four-momenta are parametrized as
	\be \label{4mompar}
		p_i^\mu  = \eta_i \omega_i \hat q^\mu (z_i, \bz_i),
	\ee
	where  $\omega_i$ is positive and real, $\eta_i = \pm 1$ for outgoing/incoming particles and 
	\be
		\begin{split}
			\hat q^\mu (z, \bz) & = \big(1+z \bz, z+\bz, -i (z-\bz), 1-z \bz\big).  
		\end{split}
	\ee
	This parametrization can be derived from the spinor helicity variables \eqref{sphelvar} presented in the introduction 
	by taking 
	\be
		p^\mu = \sigma^{\mu \dot \alpha \alpha}p_{\alpha \dot \alpha} , \quad \quad \quad
			 \sigma^{\mu \dot \alpha \alpha} = \sigma^1\big ({\bf 1}, \sigma^i \big) \sigma^1. 
	\ee
	Momenta parametrized as in \eqref{4mompar} obey
	\be
		p_i \cdot p_j =- 2 \eta_i \eta_j \omega_i \omega_j z_{ij} \bz_{ij}.
	\ee
	Polarization vectors for positive and negative helicity photons are given by
	\be \label{polpar}
		\ve_+^\mu = \frac{1}{\sqrt{2}} \p_z \hat q^\mu(z, \bz), \quad \quad \quad \ve_-^\mu = \frac{1}{\sqrt{2}} \p_\bz \hat q^\mu(z, \bz),
	\ee
	and polarization tensors for positive and negative helicity gravitons are constructed from polarization vector for photons in the following way
	\be \label{tpar}
		\ve_{\pm}^{\mu\nu} = \ve_{\pm}^\mu \ve_\pm^\nu.
	\ee
	Here we employ the following mode expansion for the photon field operator
		\be
			\hat A_\mu (x) = e \int \frac{d^3 \vec q}{(2 \pi)^3} \frac{1}{2 q^0}
				 \left[\ve_\mu^{*\alpha}  a_\alpha( \vec q) e^{i q \cdot x}+\ve_\mu^{\alpha}  a^\dagger_\alpha( \vec q) e^{-i q \cdot x}  \right]
		\ee
		and graviton field operator
		\be
			\hat h_{\mu \nu}(x) = \frac{\kappa}{2} \int \frac{d^3 \vec q}{(2 \pi)^3} \frac{1}{2 q^0}
				 \left[\ve_{\mu \nu}^{*\alpha}  a_\alpha(\vec q) e^{i q \cdot x}+\ve_{\mu \nu}^{\alpha}  a^\dagger_\alpha( \vec q) e^{-i q \cdot x}  \right].
		\ee
		The creation and annihilation modes for both obey
		\be
			\left[ a_\alpha(\vec q),  a^\dagger_\beta( \vec k)\right] = \delta_{\alpha \beta} (2 \pi)^3 (2 q^0) \delta^{(3)} (\vec q - \vec k).
		\ee
	
	Focussing on $2 \to 2$ scattering processes, we take
	\be
		\eta_1 = \eta_2= -\eta_3= -\eta_4 = -1,
	\ee
	and obtain the following expressions for the Mandelstam invariants
	\be
		\begin{split}
			s& = -(p_1 +p_2)^2 = 4 \omega_1 \omega_2 z_{12} \bz_{12}
			,\\
			t& = -(p_1 +p_3)^2 = -4 \omega_1 \omega_3 z_{13} \bz_{13}
			,\\
			u=-s-t& = -(p_2 +p_3)^2 = -4 \omega_2 \omega_3 z_{23} \bz_{23}
			.
		\end{split}
	\ee
	A $2 \to 2$ scattering amplitude of massless particles in momentum space can always be put in the general form
	\be \label{spin_gen}
		{\bf A} (p_1, p_2, p_3, p_4)
			 = H(z_i, \bz_i) \ms (s, t) \delta^{(4)} \Big(\sum_{i=1}^4 p_i\Big).
	\ee
	Here, $H$ is a kinematical factor
	\be  
		\begin{split}
		H(z_i, \bz_i) &=   \prod_{i < j}\left(\eta_i \eta_j\frac{ \langle ij \rangle }{ [ ij ]} \right)^{\frac{1}{2} (\frac{1}{3}s -  s_i - s_j)} \\
			&=   \prod_{i < j}\left(z_{ij} \bz_{ij}^{-1} \right)^{\frac{1}{2} (\frac{1}{3}s -  s_i - s_j)}  , \quad \quad\quad  s= \sum_{i=1}^4 s_i,
		\end{split}
	\ee
	which accounts for the non-trivial action of the Lorentz group on states of helicity $s_i$ massless particles.  The dynamical content of the amplitude is 
	captured by ${\bf M} (s, t)$.   For example,  a massless scalar amplitude has $H =1$,  while for $--\to ++$ graviton scattering  
	\be
		H (z_i, \bz_i) = \frac{\langle 12\rangle ^2 [34]^2}{[12]^2 \langle 34\rangle^2}=\frac{z_{12}^2\bz_{34}^2}{\bz_{12}^2 z_{34}^2}.
	\ee

	We can similarly disentangle kinematic and dynamic contributions to a generic  4-point celestial amplitude (suppressing helicity labels)
	\be
		\ca(\Delta_i, z_i, \bz_i) = \Big(\prod_{i = 1}^4 \int_0^\infty \frac{d\omega_i}{\omega_i}~ \omega_i^{\Delta_i}  \Big) {\bf A} (p_1, p_2, p_3, p_4) 
	\ee
	by making a change of variables from  $\omega_i$ to $(\omega, \sigma_2, \sigma_3, \sigma_4)$ where
	\be
		s = \omega^2, \quad \quad \quad \sigma_i = \omega_i/\omega.
	\ee
	The measure transforms as 
	\be
		\prod_{i = 1}^4 \int_0^\infty \frac{d \omega_i}{\omega_i}=2 \int_0^\infty \frac{d \omega }{\omega} \prod_{i = 2}^4 \int_0^\infty \frac{d \sigma_i}{\sigma_i},
	\ee
	and the momentum-conserving delta function included in ${\bf A}$ can be put in the form 
	\be \label{deltafunc}
		\begin{split}
			\delta^{(4)}\bigg(\sum_{i =1}^4 p_i\bigg)
								& =\frac{\delta (z-\bz)}{4 \omega^4}\sqrt{\frac{z (1-z)}{z_{23} \bz_{23} z_{24} \bz_{24} z_{34} \bz_{34}}} 
								 \delta \bigg (\sigma_2 -\frac{1}{2} \sqrt{ \frac{z (1-z)z_{34} \bz_{34}}{z_{23} \bz_{23} z_{24} \bz_{24}}}   \bigg)\\& \quad \quad \times
									 \delta  \bigg (\sigma_3-\frac{1}{2}\sqrt{\frac{1-z}{z}\frac{z_{24} \bz_{24}}{ z_{34} \bz_{34} z_{23} \bz_{23}}}  \bigg ) 
									 \delta \bigg (\sigma_4 -\frac{1}{2}\sqrt{\frac{z}{1-z} \frac{z_{23} \bz_{23}}{z_{24}\bz_{24}z_{34}\bz_{34}}}  \bigg ),
		\end{split}
	\ee
	where  the cross ratio $z$ defined in \eqref{crossratio} lies in the range $0<z<1$ for this channel.
	
	On the support of the delta function, in these new variables, the Mandelstam invariants simplify to
	\be
		\begin{split}
			s = \omega^2, \quad \quad \quad t= -z \omega^2, \quad \quad \quad u = -(1-z) \omega^2.
		\end{split}
	\ee
	Finally, performing the $\sigma_i$ integrals, the celestial 4-point amplitude can be written in the form 
	\be
		\ca (\Delta_i, z_i, \bz_i)  = X \mathcal{A}(\beta, z),
	\ee
	where $X$ is the kinematic factor
	\be \label{defX}
		X =  \Big( \prod_{i < j} z_{ij}^{\frac{h}{3} - h_i - h_j} \bz_{ij}^{\frac{\bar{h}}{3} - \bar{h}_i - \bar{h}_j} \Big) \delta (z-\bz)
					 2^{-\beta -2}     
								   		\left|z(1-z)\right|^{\frac{1}{6}(\beta +4)}.
	\ee
	In the above, $h_i$ ($\bh_i$) are the standard left and right conformal weights
	\be
		h_i = \frac{1}{2} (\Delta_i+ s_i), \quad \quad \quad \bh_i = \frac{1}{2} (\Delta_i- s_i),  
	\ee
	and their sums are denoted by $h$ and $\bh $, respectively.
	$\mathcal{A}(\beta,z)$ is explicitly related to the  matrix element $\ms$ by
	\be
		\begin{split}
			 \mathcal{A}(\beta, z) &=   \int_0^\infty \frac{d \omega }{\omega} \omega^{\beta} \ms(\omega^2, -z \omega^2),
		\end{split}
	\ee 
	and therefore represents the dynamical content of four-point celestial amplitudes. 
	
	Writing the  $-- \to ++$ momentum space   four-graviton scattering amplitude in Einstein gravity at tree-level in the form 
	\be
		{\bf A}(p_i) = - \frac{\langle 12 \rangle^2[34]^2}{[12]^2 \langle 34\rangle^2} \frac{G s^3}{t(t+s)} \delta^{(4)} \Big(\sum_{k=1}^4 p_k\Big),
	\ee
	we readily  obtain its celestial representation
	\be
		\ca(\Delta_i, z_i, \bz_i) = X   \int_0^\infty \frac{d \omega }{\omega} \omega^{\beta} \frac{G \omega^2 }{z (1-z)}.
	\ee

\end{document}